\documentclass[11pt,a4paper]{article}
\pdfoutput=1
\usepackage{jheppub}

\usepackage{ulem}
\usepackage{epsfig}
\usepackage{latexsym}
\usepackage{amsfonts}
\usepackage{amsmath}
\usepackage{amsthm}
\usepackage{amssymb}
\usepackage{amsbsy}
\usepackage{multirow}

\newcommand{\CC}{\mathbb{C}} 
\newcommand{\RR}{\mathbb{R}} 
\newcommand{\ZZ}{\mathbb{Z}} 

\def\tr         {{\rm  tr}}

\def\calc         {{\cal C}}

\def\calm         {{\cal M}}

\def\calo         {{\cal O}}
\def\calp         {{\cal P}}

\def\calw         {{\cal W}}

\def\cC{{\cal C}}
\def\cD{{\cal D}}

\def\cG{{\cal G}}

\def\cR{{\cal R}}

\newsavebox{\uuunit}
\sbox{\uuunit}
    {\setlength{\unitlength}{0.825em}
     \begin{picture}(0.6,0.7)
        \thinlines
        \put(0,0){\line(1,0){0.5}}
        \put(0.15,0){\line(0,1){0.7}}
        \put(0.35,0){\line(0,1){0.8}}
       \multiput(0.3,0.8)(-0.04,-0.02){12}{\rule{0.5pt}{0.5pt}}
     \end {picture}}
\newcommand {\unity}{\mathord{\!\usebox{\uuunit}}}

\def\be{\begin{equation}}
\def\ee{\end{equation}}
\def\bea{\begin{eqnarray}}
\def\eea{\end{eqnarray}}

\def\a{\alpha}
\def\b{\beta}

\def\d{\delta}
\def\e{\epsilon}
\def\D{\Delta}
\def\l{\lambda}
\def\L{\Lambda}

\def\f{\phi}

\def\m{\mu}
\def\n{\nu}
\def\o{\omega}

\def\p{\pi}
\def\r{\rho}

\def\s{\sigma}
\def\S{\Sigma}

\def\sF{{{ F}\!\!\!\!\hskip.8pt\hbox{\raise1pt\hbox{/}}\,}}
\def\som{{{ \omega}\!\!\!\!\hskip.8pt\hbox{\raise1pt\hbox{/}}\,}}
\def\sJ{{{\rm J}\!\!\!\!\hskip.8pt\hbox{\raise1pt\hbox{/}}\,}}

\def\pa{\partial}

\def\to{\rightarrow}
\def\nonu{\nonumber \\{}}
\def\half{\frac{1}{2}}

\theoremstyle{definition}
\newtheorem*{remark}{Remark}
\newcommand{\rem}{\begin{remark}}
\newcommand{\erem}{\end{remark}}

\title{Quantizing higher-spin gravity in free-field variables}
\author[a]{Andrea Campoleoni,}
\author[b,c]{Stefan Fredenhagen}
\author[d]{and Joris Raeymaekers}

\affiliation[a]{Institut f\"ur Theoretische Physik, ETH Zurich,\\
Wolfgang-Pauli-Strasse 27, 8093 Z\"urich, Switzerland}
\affiliation[b]{University of Vienna, Faculty of Physics,\\ Boltzmanngasse 5, 1090 Vienna, Austria}
\affiliation[c]{Erwin Schr{\"o}dinger International Institute for Mathematics and Physics,\\ University of Vienna, Boltzmanngasse 9, 1090 Vienna, Austria}
\affiliation[d]{CEICO, Institute of Physics of the ASCR,\\  Na Slovance 2, 182 21 Prague 8, Czech Republic}

\emailAdd{campoleoni@itp.phys.ethz.ch}
\emailAdd{stefan.fredenhagen@univie.ac.at}
\emailAdd{joris@fzu.cz}

\abstract{We study the formulation of  massless higher-spin gravity on AdS$_3$ in  a gauge in which
the fundamental variables satisfy free field Poisson brackets. This gauge choice  leaves a small portion of the gauge freedom  unfixed, which  should be further quotiented out. We show that doing so leads to a  bulk version of the  Coulomb gas formalism for $W_N$ CFT's: the generators of the residual gauge symmetries are  the classical limits of  screening charges, while the gauge-invariant observables are  classical $W_N$ charges. 

Quantization in these variables can be carried out using standard techniques and makes manifest a remnant of the  triality symmetry of $W_\infty[\l]$.
  This symmetry can be used to argue that the theory should be supplemented with 
 additional matter content  which is precisely that  of the Prokushkin-Vasiliev theory.
As a further  application, we use our formulation to quantize a class of  conical surplus solutions 
and confirm the conjecture that these  are  dual to specific degenerate $W_N$ primaries, to all orders in the large central charge expansion.}
\arxivnumber{}
\preprint{UWThPh-2017-44}
\begin{document}
\maketitle

\section{Introduction}

The AdS/CFT correspondence offers a nonperturbative definition of quantum gravity in anti-de Sitter space
in the form of  a dual conformal field theory (CFT), thus making conceptual problems in quantum gravity accessible, at least in principle.
Being  a strong-weak coupling duality, a complete proof of the correspondence may 
be out of reach in the  richest  examples, though one may hope to do better in the recently uncovered simpler and more tractable instances of holographic duality.
One promising class of dualities was identified by Gaberdiel and Gopakumar \cite{Gaberdiel:2010pz}, who proposed that a $(2+1)$-dimensional
higher-spin gravity theory due to Prokushkin and Vasiliev \cite{Prokushkin:1998bq} is  holographically dual to a certain large $N$ limit of the $W_N$ minimal model CFT's. 
The bulk theory  allows for a consistent truncation to the  massless higher-spin sector which does not contain any locally propagating fields and can be elegantly described as a Chern-Simons theory. The only degrees of freedom in this sector are boundary fluctuations, which are furthermore controlled by a large $W_\infty[\lambda]$ symmetry.

Far from giving a complete proof of the higher-spin/minimal-model duality, we will in this work
study  a set of variables for the massless higher-spin sector in which quantization is relatively straightforward; these variables can roughly be  thought of as Darboux coordinates on the phase space of higher-spin fields.
 For simplicity, we restrict our attention to the theory with higher-spin algebra $sl(N, \CC)$ which describes spins $2, \ldots, N$ in Euclidean signature. This theory is based on an $sl(N, \CC)$ Chern-Simons gauge field
 with  boundary conditions which impose asymptotic AdS behavior of the field. These boundary conditions
 are preserved by a   group of `Drinfeld-Sokolov (DS) gauge symmetries', and quotienting these out
 (`DS reduction' \cite{Drinfeld:1984qv})  leads to a theory with classical $W_N$ symmetry \cite{Henneaux:2010xg,Campoleoni:2010zq}. 
 
 In CFT's  with quantum $W_N$ symmetry, the Coulomb gas formalism \cite{Fateev:1987zh} (see also the review \cite{Lukyanov:1990tf}) gives a realization of  the $W_N$ currents as operators built out of $N-1$ free fields which  commute with a set of screening operators.
On the bulk dual side, it has also been known since \cite{Balog:1990mu} that there exists a gauge for the
DS gauge symmetry of the higher-spin theory, the `diagonal gauge', in which the Poisson brackets of the fundamental variables reduce to those of free fields. However it was also pointed out in \cite{Balog:1990mu} that this gauge choice suffers from  ambiguities in  that there are  many diagonal gauge representations of the same physical configuration.  Also, it was not clear how the screening charges, which are vital for obtaining a $W_N$ symmetry, appear in this gauge, and if they are related to the aforementioned ambiguity.

In this work, we will address these puzzles in detail.  We will show that the ambiguities observed in 
\cite{Balog:1990mu}  arise from the fact that the diagonal gauge does not fix the DS gauge symmetry completely; rather it leaves a finite-parameter family of  residual DS gauge transformations. These should  be viewed as proper gauge transformations which do not change the physical state.\footnote{This is the main difference with respect to the formally similar gauge choice of  \cite{soft-hairs,HS-soft-hairs}. In the latter context, the boundary conditions have been designed such that all residual symmetries are actual global symmetries, thus leading e.g.\ to a theory with a different asymptotic symmetry algebra with respect to that obtained with a DS reduction.}
We will see that the charges which generate these residual DS transformations through Poisson brackets are precisely the classical limits of the screening charges of the Coulomb gas formalism for $W_N$ CFT's. In addition, the observables which are invariant under the residual DS  gauge symmetry are precisely the classical $W_N$ currents. Therefore all the ingredients of the Coulomb gas formalism appear naturally when considering bulk higher-spin theory in the diagonal gauge.

Once the theory is formulated in free field variables, quantization can be carried out using standard techniques, and we will see
that it leads to a quantum shift in the background charge of the free fields.  This uncovers 
a $\ZZ_2$ duality symmetry which is not visible at the classical level and is a remnant of the triality symmetry of the $W_\infty[\lambda]$ algebra \cite{Gaberdiel:2012ku}. This symmetry can be used to argue, modulo some natural assumptions, that the theory should be supplemented with 
 additional matter content  which is precisely that  of the Prokushkin-Vasiliev theory.

When investigating the effect of the residual DS gauge transformation, we find a qualitative difference depending on whether we consider  `generic' field configurations, in which the gauge field is allowed to have isolated singularities around which the holonomy is nontrivial, or the `non-generic' case where  the holonomy is trivial everywhere. In the former case, the residual DS gauge transformations are discrete, giving a realization of the symmetric group, while in the latter case they  act as continuous transformations.  
To illustrate the power of the free field formulation, we address in detail  the quantization 
of the latter class of configurations with trivial holonomy. These were considered before in \cite{Castro:2011iw} 
and were shown to describe  a class of generalized  conical surpluses, and  in our   variables they appear as winding modes of the free fields. These solutions were conjectured to be dual to a subclass of degenerate primaries of 
the quantum $W_N$ algebra in \cite{Castro:2011iw,Perlmutter:2012ds}, and evidence for this has accumulated  \cite{Campoleoni:2013iha,Raeymaekers:2014kea}.\footnote{The same states have been also recently considered in a related holographic scenario, where they have been identified in the twisted sector of symmetric orbifold CFT's, which have been conjectured to be dual to the tensionless limit of strings on AdS$_3$ \cite{Datta:2016cmw}.} Upon quantizing these solutions in our
framework, we  are able to confirm  this conjecture to all orders in the $1/c$ expansion.

\section{Review of higher-spin gravity in  Chern-Simons formulation}\label{sec:review}

In this section we recall the key features of the Chern-Simons formulation of Euclidean higher-spin gauge theories in three-dimensional anti-de-Sitter (AdS) background. In particular, we review the boundary conditions that select asymptotically-AdS field configurations, stressing that they can be formulated in many different, albeit equivalent, ways. We focus on boundary conditions belonging to the class of Drinfeld-Sokolov gauges and we confront them with the diagonal gauge that will be analyzed in the following sections.  

\subsection{Chern-Simons theory with boundary}

Higher-spin gauge theories in three-dimensional anti-de-Sitter background, in the absence of matter couplings, can be formulated as pure Chern-Simons gauge theories \cite{Blencowe:1988gj}. Here we focus on Euclidean theories
involving spins $2,3,\dots, N$, which are described by a gauge field $A$ taking values in the complex Lie algebra $sl(N,\CC)$. For a large part of this work, where we consider the infinitesimal symmetries of the theory, the precise global structure of the gauge group will be immaterial. When global considerations will enter the discussion, notably in section \ref{secfinite}, we will assume the gauge group to be $SL(N,\CC)/\ZZ_N$.
This means that, as in \cite{Castro:2011iw}, we mod out by the center $\ZZ_N$, generated by the element $e^{2\p i / N} \unity$ in the defining representation, which is invisible to adjoint fields. 

 The Euclidean action is
\begin{equation}
I_E=-\,i\left( S_{CS}[A] - S_{CS}[ \bar A]\right) = 2\, {\rm Im} \left( S_{CS}[A]\right) , \label{SE}
\end{equation}
where $S_{CS}[A]$ is the Chern-Simons action
\begin{equation}
S_{CS}[A] = \frac{k}{4\p} \int_\calm \tr_{\bf N}\left( A \wedge dA + \frac{2}{3} A\wedge A\wedge A \right) .
\end{equation}
In the formulae above, $k$ is a real parameter, $\tr_{\bf N}$ is the trace in the defining $N$-dimensional representation of $sl(N,\CC)$, and $\bar A$ is the complex conjugate\footnote{Another form of the Euclidean action, which often appears in the literature, is given by \mbox{$I_E=-i( S_{CS}[A] - S_{CS}[ - A^\dagger])$}, where $\,\dagger$ stands for Hermitean conjugation. 
This form  is equivalent to ours, since one can easily show that $S_{CS}[ - A^\dagger]= S_{CS}[ \bar A]$.}
of $A$.
See \cite{Witten:1989ip} for more details on the Chern-Simons formulation of Euclidean gravity and e.g.\ Appendix A.1 of \cite{Bunster:2014mua} for its higher-spin extensions. We only consider the case in which the spin-$2$ subsector corresponds to the principal embedding of $sl(2,\CC)$ into  $sl(N,\CC)$. As shown in Appendix \ref{AppNewton}, with this choice the constant $k$ is related to the AdS$_3$ radius $l$ and the 3D Newton constant $G_N$ as
\begin{equation}
k = - \frac{3 l}{2 N(N^2-1) G_N}\,.\label{kitoGN}
\end{equation}
Note that, in our conventions, the weakly coupled gravity regime corresponds to taking $k$ large and negative.

To simplify the comparison of our ensuing analysis with CFT techniques, we assume the manifold $\calm$ to have the topology of a solid sphere. The boundary $S^2$ corresponds to the usual cylindrical boundary of $AdS_3$ where we added the two points at infinity. In actual computations, following the most common CFT setup, we will perform an additional conformal transformation and describe the boundary as a plane parameterized by a complex coordinate $z$. In these conventions, Euclidean $AdS_3$ can be parameterized e.g.\ as $ds^2 = d\r^2 + e^{2\r}dz d\bar{z}$, where $\r$ is a radial coordinate transverse to the boundary, while slices of constant Euclidean time are circles of constant $|z|$.

Upon varying the action \eqref{SE} one picks up a boundary term:
\begin{equation}
\d S_{CS} = - \frac{k}{4\p} \int_{\partial \calm} \tr_{\bf N} \left( A \wedge \d A \right) +  {\rm (eom)} \, .
\end{equation}
To have a good variational principle, we make it vanish by imposing the boundary conditions
\begin{equation}
(A_{\bar z})|_{ \partial \calm} = 0 \,, \qquad (\bar A_{ z})|_{ \partial \calm} =0 \,. \label{CSboundcond}
\end{equation}
For $N > 2$ these constraints are too restrictive to accommodate black holes \cite{GK_black-hole}, but they still allow to identify the asymptotic symmetries of the full space of solutions of the equations of motion \cite{Henneaux:2010xg,Campoleoni:2010zq,Bunster:2014mua}. Moreover, they are satisfied by the solutions we will discuss in section \ref{seccondef}, so that we stick to this simplifying choice. Actually, we further constrain the gauge connections to be of the form
\begin{equation}
A = b^{-1} a_z (z) b\, dz + b^{-1} d b \,, \qquad \bar A = b^{-1} \bar a_{\bar z}(\bar z) b\, d\bar z + b^{-1} d b \,, \label{gfcs}
\end{equation}
where $b = b(\r)$ is a group element depending only on the coordinate $\r$ transverse to the boundary. As discussed e.g.\ in \cite{Banados:1998gg,Campoleoni:2010zq}, this form of the connection can be reached by combining a residual gauge fixing with the equations of motion.

Concretely, we take $b = e^{\r J_0}$, where $J_0$ is a generator of the  $sl(2)$ subalgebra which is  principally embedded in $sl(N)$. For later convenience, we give an explicit representation for the generators of this $sl(2)$:
\begin{equation}
J_0 = \sum_{i=1}^N \left( \frac{N+1}{2} - i\right) e_{ii}\,,\qquad
J_{1} = - \sum_{i = 1}^{N-1} e_{i+1,i}\,,\qquad
J_{-1} = \sum_{i = 1}^{N-1} i (N-i)e_{i, i+1}\,, \label{defJ}
\end{equation}
where the $e_{ij}$ are $N$-dimensional matrices with entries $(e_{ij})_k{}^l = \d_{ik} \d_j{}^l$. These represent the algebra $ [J_m,J_n] =(m-n) J_{m+n}$.

The gauge choice (\ref{gfcs}) is invariant under residual `gauge' transformations with parameters $\L = b^{-1} \l (z)  b$ and $\bar \L =  b^{-1} \bar \l (\bar z)  b $, which induce infinitesimal shifts of $a_z$ and $\bar{a}_{\bar{z}}$ of the form
\begin{equation}
\d a_z = \pa_z \l + [a_z,\l ] \, , \qquad \d \bar a_{\bar z} = \pa_{\bar z}\bar \l + [\bar a_{\bar z},\bar \l ] \, ,\label{gaugetransf}
\end{equation}
where $\l (z) \in sl(N)$  
is a purely holomorphic  Lie algebra element. These transformations must actually be interpreted as global symmetries \cite{Banados:1994tn}, whose associated conserved charges can be derived as follows (see e.g.\ \cite{Banados:1998gg,Campoleoni:2010zq} for more details). The canonical generator of gauge transformations in the Chern-Simons theory is
\begin{equation}
G(\L) = \frac{k}{4\p} \int_\S dx^i \wedge dx^j \, \tr_{\bf N} \left( \L F_{ij} \right) + Q(\L) \, ,
\end{equation}
where $F_{\m\n} = \pa_\mu A_\nu -\pa_\nu A_\mu +[A_\mu,A_\nu]$ is the field strength, while the $x^i$ are the coordinates on a slice of fixed time that we denoted by $\S$. $Q(\L)$ is a boundary term, whose role is to cancel the boundary contribution produced by the functional variation of $G(\L)$ that determines its Poisson bracket with the fields:
\begin{equation} \label{varQ}
\d G(\L) = - \frac{k}{2\p} \int_\S dx^i \wedge dx^j\, \tr_{\bf N} \big( (\pa_i\L+[A_i,\L])   \d A_j \big) + \left( \frac{k}{2\p} \int_{\partial \S} dx^i \,\tr_{\bf N} \left( \L \d A_i \right) + \d Q(\L) \right) .
\end{equation}
For field-independent gauge parameters one can readily integrate the variation $\d Q(\L)$ that guarantees the cancellation of the terms between brackets. In our conventions, one can rewrite the ensuing charges as contour integrals in the $z$-plane:
\be
Q (\l, \bar \l) = - \frac{k}{2 \p}  \oint \tr_{\bf N} \left( \l\, a_z dz +  \bar \l\, \bar a_{\bar z} d\bar z\right) . \label{consQ2}
\ee
The integration contour is originally a counterclockwise circle of constant Euclidean time $|z|$, which of course can be deformed as long as no singularities of the integrand are encountered.

The charges \eqref{consQ2} play the dual role of being conserved through the Noether theorem, but also of generating the associated global symmetries through the Dirac bracket as
\begin{equation}
    \delta a_z = \{ Q(\lambda,\bar{\lambda}) , a_z\}\label{DBdef}
\end{equation}
and its analogue for $\bar{a}_{\bar{z}}$ \cite{Banados:1994tn}. This relation actually provides an efficient way to compute the Dirac bracket of the fields.
Expanding them as
\begin{equation} \label{modes-a}
    a_z(z) = \sum_{m\in\mathbb{Z}} a_m \,z^{-m-1} \, ,\qquad \bar{a}_{\bar{z}}  = \sum_{m\in\mathbb{Z}} \bar{a}_m \, \bar{z}^{-m-1} \, 
\end{equation}
and introducing arbitrary elements $\alpha,\beta$ of $sl(N)$, by substituting \eqref{gaugetransf} and \eqref{consQ2} in \eqref{DBdef} one obtains the affine $sl(N)$ Lie algebra
\begin{equation}\label{affineLie}
    \{ \tr_{\bf N} ( \alpha\, a_m) , \tr_{\bf N}(\beta\, a_n) \} = \frac{i}{k} \, \Big( \tr_{\bf N} ([\alpha,\beta]\, a_{m+n}) + m \,\tr_{\bf N} (\alpha \beta ) \delta_{m+n,0} \Big) \, 
\end{equation}
together with its analogue for $\bar{a}_{\bar{z}}$. 

\subsection{AdS asymptotic conditions and Drinfeld-Sokolov gauges}\label{secadsas}

Selecting AdS asymptotic behavior requires additional constraints on the strictly lower-triangular part $(a_z)_+$ of the gauge potential $a_z$ \cite{Banados:1998gg,Henneaux:2010xg,Campoleoni:2010zq}:
\be
(a_z)_+  \approx J_1 \, ,\label{asads}
\ee
where $J_1$ is one of the generators of the $sl(2)$ subalgebra defined in~\eqref{defJ}. The symbol $\approx$ stresses that we are imposing a constraint on the phase space of the boundary degrees of freedom of the Chern-Simons theory with boundary conditions~\eqref{CSboundcond}. This is a first class constraint: indeed it can be formulated as 
\begin{equation}\label{constr_alpha}
    \tr_{\bf N} \big(\alpha_{-} (a_m - J_1 \delta_{m,-1})\big) \approx 0
\end{equation}
for all strictly upper-triangular (constant) matrices $\alpha_-$, which gives
\begin{equation}
    \{\tr_{\bf N} \big(\alpha_{-} (a_m - J_1 \delta_{m,-1})\big) , \tr_{\bf N} \big(\beta_{-} (a_n - J_1 \delta_{n,-1})\big) \} = \frac{i}{k}\, \tr_{\bf N} \big([\alpha_{-},\beta_{-}] a_{m+n}\big)
\end{equation}
thanks to~\eqref{affineLie}.
The right-hand side is again proportional to the constraint because the commutator of two strictly upper-triangular matrices has only entries starting from the second upper diagonal, and therefore
\begin{equation}
    \tr_{\bf N} \big([\alpha_{-},\beta_{-}] a_{m+n}\big) = \tr_{\bf N} \big([\alpha_{-},\beta_{-}] (a_{m+n} - J_1 \delta_{m+n,-1})\big) \, .
\end{equation}
This result allows one to interpret the
transformations (\ref{gaugetransf}) preserving the constraint \eqref{asads}, which are generated by arbitrary strictly upper-triangular matrices
\be \l (z) = \l_- (z) \, , \label{uppertriang}\ee  
as a proper gauge symmetry.

The upper-triangular gauge redundancy (\ref{uppertriang}) can be eliminated by imposing suitable gauge-fixing conditions. Most prevalent in the literature are those belonging to the class of Drinfeld-Sokolov gauges \cite{Drinfeld:1984qv,Balog:1990mu}. These fix the residual gauge symmetry completely, so that the full set of constraints becomes second class.
Let us illustrate the situation in the case of $sl(2)$. We expand the modes $a_m$ of $a_z$ in the generators $J_i$ of $sl(2)$,
\begin{equation}
    a_m = \sum_{i=-1}^1 a_m^i\, J_i\, .
\end{equation}
The constraint~\eqref{asads} then reads
\begin{equation}
    \mathcal{C}_m = a^1_m -\d _{m,-1} \approx 0 \, ,
\end{equation}
and the (in this case unique) Drinfeld-Sokolov gauge-fixing condition is
\begin{equation}
    \cG_m = a^{0}_m \approx 0 \, .
\end{equation}
The constraints satisfy the Poisson brackets
\begin{subequations} \label{constr-DS}
\begin{align}
    \{\cC_m , \cG_n \} &= -\frac{2 i}{k}\, a^1_{m+n} \approx  -\frac{2 i}{k}\,  \d_{m+n,-1}\, , \\
    \{\cG_m , \cG_n \} &= \frac{2i\,m}{k}\, \d_{m+n,0} \, , \\ 
    \{\cC_m , \cC_n \} &= 0 \, ,
\end{align}
\end{subequations}
so that the constraint matrix is invertible. The constraints are therefore second class, which confirms that $\cG_m$ is a complete gauge fixing. 

In this work we consider another gauge which is outside of the Drinfeld-Sokolov class. In the $sl(2)$ case it is specified by the partial gauge-fixing condition
\begin{equation}
    \cD_m = a^{-1}_m \approx 0 \, . \label{diagsl2}
\end{equation}
The constraint matrix has entries
\begin{subequations} \label{constr-diag}
\begin{align}
    \{\cC_m , \cD_n \} &= - \frac{i}{k}\left( a^0_{m+n} + m\, \d_{m+n,0} \right) , \\
    \{\cD_m , \cD_n \} &= 0 \, , \\
    \{\cC_m , \cC_n \} &= 0 \, ,
\end{align}
\end{subequations}
and it is not invertible on the whole constraint surface due to the dependence on $a^0$. 
An example of a configuration on which invertibility fails is when
\be 
a^0_m =  -\, \L\, \d_m^0 \, , \quad \textrm{with}\ \L \in \ZZ \, . \label{cdfirst}
\ee
One sees from (\ref{constr-diag}) that on this configuration, $\cC_\L $ and
$\cD_{-\L} $ remain as first class constraints. These configurations and their $sl(N)$ generalizations   will play an important role in what follows.

The generalization of the  partial gauge-fixing (\ref{diagsl2}) to $sl(N)$ amounts to constraining the strictly upper-triangular part of the fields $a_z$ to vanish, i.e. 
\begin{equation}
    (a_z)_-\approx 0 \, ,\label{diagslN}
\end{equation}
which we will refer to as the `diagonal gauge'. The subtleties associated to this gauge choice will be discussed in the next two sections.

\section{The diagonal gauge}\label{sec:diagonalgauge}

In this section we study the formulation of higher-spin gravity in the diagonal gauge (\ref{diagslN}).
As we shall see, this gauge  has the great advantage that the Poisson brackets of the basic fields become extremely simple:  they are those of free fields. On the other hand,
it is not a Drinfeld-Sokolov gauge and this leads to some puzzles which were first anticipated in \cite{Balog:1990mu} and which we aim to address in this work. They  
stem from the property we already noted above in the $N=2$ case, namely that the diagonal gauge does not completely fix the proper gauge freedom (\ref{uppertriang}). After discussing the origin of these  residual gauge symmetries, we  study
the corresponding gauge orbits in detail. This paves the way for the discussion, in the next section, of how to factor out the residual symmetries in order
to obtain a theory equivalent to the one  formulated in a Drinfeld-Sokolov gauge.

\subsection{Reaching the diagonal gauge}

In the diagonal gauge  the nontrivial components of the connection reside on the main diagonal, so that it can be parameterized as follows: 
\be
a_z= J_1 + \frac{1}{\sqrt{k}}\sum_{j=1}^N  \vec{\e}_j \cdot (i \pa_z \vec{\phi})\, e_{jj} \,. \label{diaggauge}
\ee
Here, $\pa_z \f^a (z)$, with $a= 1,\ldots , N-1$,  form a  vector of purely $z$-dependent fields (the arrow notation $\vec v$ will be used to denote  $(N-1)$-component vectors). 
 We will impose also some regularity conditions on the fields $\pa_z \f^a (z)$, which we shall spell out when they become relevant for the discussion, namely in section \ref{secfinite}.   The set of  $N$ vectors
$\vec{\e}_i$, with $i=1,\ldots , N$, are instead the weights of the vector representation of $sl(N)$. They  satisfy the properties (see e.g.\ section 6.3.3 of \cite{Bouwknegt:1992wg})
 \be
 \vec{\e}_i \cdot \vec{\e}_j = \d_{ij} - \frac{1}{N} \,, \qquad
 \sum_{i=1}^N \vec \e_i = 0 \,,\qquad \sum_i  \e_{ia} \e_{i b} = \d_{ab} \,.\label{epsprops}
 \ee
We recall for later use the relation between the $\vec{\e}_i$ and simple roots $\vec e_a$ and fundamental weights $\vec \o_a$ of $sl(N)$:
\begin{align}
\vec \o_a &= \sum_{i=1}^a \vec \e_i\, , &
 \vec \e_i &=  \vec \o_i - \vec \o_{i-1}\, , &
\vec e_a &=  \vec{\e}_a - \vec{\e}_{a+1}\, . \label{epsprops2}
\end{align}
According to (\ref{epsprops}), the roots are normalized to have length squared equal to 2.

By taking the complex conjugate of (\ref{diaggauge}) one obtains an analogous parameterization of  $\bar a_{\bar z}$  in terms of the antiholomorphic fields $\pa_{\bar z} \bar \f^a (\bar z)$; we will usually display only the expressions  for the holomorphic sector in what follows.

Locally, the gauge (\ref{diaggauge}) is reachable by making a finite proper gauge transformation of the type (\ref{uppertriang}) as for a Drinfeld-Sokolov gauge. To see this, we start from an arbitrary flat connection
$a_z$ satisfying the constraint (\ref{asads}). We note that locally we can  write the gauge  potential in pure gauge form, $a_z = g^{-1} \pa_z g$, with $g(z)$ a $z$-dependent element of the gauge group. Note that
 $g$ is determined only up to a left multiplication by a constant group element 
\be 
g(z) \to \tilde \L_0 g(z)\,.
\ee
Finite gauge transformations arise from right multiplication by a $z$-dependent group element
\be 
g(z) \to g(z) \L (z)\,.
\ee
We now make a Gauss decomposition 
\be 
g = N_+ h N_-\,,\label{Gaussdec}
\ee
where $h$ is a diagonal group element and $N_+$ and $N_-$ are  lower and upper triangular group elements, respectively, whose diagonal entries are all equal to one.
To reach the diagonal gauge, we make a finite upper-triangular gauge transformation with parameter $\L =N_-^{-1}$ so that
\be 
g = N_+ h \, .\label{puregauge}
\ee
In this gauge, the  asymptotic AdS condition (\ref{asads}) is satisfied for arbitrary $h$ and for $N_+$ determined in terms of $h$, up to multiplication from the left by a constant matrix $\tilde{\L}_0$,  by the equations
\be
N_+^{-1} \pa_z N_+ = h J_1 h^{-1}\,.\label{Npitoh}
\ee
The form   (\ref{diaggauge}) then arises from choosing the following parameterization of $h$:
 \be
h (z) =  \sum_{j=1}^N  \exp \left({  \frac{1}{\sqrt{k}}\,   \vec{\e}_j \cdot (i \vec{\phi}(z))  } \right)  e_{jj} \,.\label{gdiag}
 \ee

We should stress that the above discussion was purely local, and that the group element $g(z)$ in (\ref{puregauge}) is not guaranteed to be globally defined. We will discuss the conditions under which
$g(z)$  is well-defined and their significance in section \ref{secfinite}.

\subsection{Residual gauge symmetries}\label{sec:residualgaugesymmetries}

Though the above argument shows that  we can  always locally reach the diagonal gauge (\ref{diaggauge}), the resulting representation turns out not to be unique, as was  observed in \cite{Balog:1990mu}. 
The origin of this non-uniqueness lies in the fact that the diagonal gauge does not completely fix the proper upper triangular gauge freedom
(\ref{uppertriang}) but allows for a small (generated by a finite number of constant parameters) amount of  residual gauge symmetry. Let us establish the existence of these  symmetries first from the point of view of the gauge potential.

Given $a_z=J_1 + \cD$ in diagonal gauge ($\cD$ diagonal), we look for the proper gauge transformations, i.e.\  for matrices $\L_-(z)= \unity + \lambda_- (z)$ with $\lambda_-$ a strictly upper triangular matrix, such that
\begin{equation}
    \tilde{a}_z = \L_-^{-1}\partial \L_- + \L_-^{-1}\,a_z\, \L_- = J_1 + \tilde{\cD} \, ,
\end{equation}
where here and hereafter we denote $\partial_z$ simply by $\partial$, and $\tilde{\cD}$ is diagonal.
This is equivalent to
\begin{equation}
    \tilde{\cD} = \L_-^{-1}\partial \L_- + (\L_-^{-1}\,J_1 \,\L_- - J_1) + \L_-^{-1}\,\cD\,\L_- \, .
\end{equation}
All three terms on the right hand side are upper triangular, $\L_-^{-1}\partial \L_-$ is even strictly upper triangular. The condition that the right hand side is diagonal is therefore equivalent to
\begin{equation}
    \L_-^{-1}\partial \L_- + \L_-^{-1} \big( [J_1,\L_-]+\mathcal{D}\L_-\big)- \Big( 
    \L_-^{-1} \big( [J_1,\L_-]+\mathcal{D}\L_-\big)
    \Big)\Big|_{\text{diag}}=0\, ,
\end{equation}
where $|_{\text{diag}}$ denotes the projection to the diagonal part. Multiplying this equation by $\L_-$ from the left one arrives\footnote{using that $\big(\L_-^{-1} \,A\big)\big|_{\text{diag}} = A_{\text{diag}}=\big( A\,\L_-\big)\big|_{\text{diag}} $ for upper triangular $A$} at the result
\begin{equation}\label{condforg}
    \partial \L_- + [a_z,\L_-] - \L_-\,\Big([J_1,\L_-]\big|_{\text{diag}}\Big) = 0\, .
\end{equation}
This is a system of nonlinear (due to the last term) differential equations.  For simplicity let us discuss the case $N=2$ in some detail, where
\begin{equation}
    a_z =\begin{pmatrix} \frac{1}{\sqrt{k}}\,i \e_1 \partial \f & 0\\ -1 & \frac{1}{\sqrt{k}}\,i \e_2 \partial \f \end{pmatrix}\, , \qquad\qquad \L_- = \begin{pmatrix} 1 & \lambda_1 \\ 0 & 1\end{pmatrix} \, .
\end{equation}
The condition~\eqref{condforg} leads to the equation
\begin{equation}
    \partial\lambda_1 + \frac{i}{\sqrt{k}}\, e_1 \partial \f\, \lambda_1 + \lambda_1^2 = 0\, ,
\end{equation}
where, in agreement with \eqref{epsprops2}, $e_1 = \e_1 - \e_2$.
This is a special form of the Riccati equation (or an instance of Bernoulli's equation) which can be solved by introducing $1/\lambda_1$ as a new variable, which turns it into a linear equation. The most general non-trivial solution is
\begin{equation}
    \lambda_1 = \partial \log n_{1} \, ,
\end{equation}
where $n_{1}$ is any integral of $-e^{-\frac{i}{\sqrt{k}}\,e_1 \f}$, that is
\begin{equation}\label{condforF1}
    \partial n_{1} = -e^{-\frac{i}{\sqrt{k}}\,e_1 \f}\, .
\end{equation}
The corresponding finite proper gauge transformation is given by
\begin{equation}\label{largegaugetransfoN2}
    i\partial\f \to i\partial\f + \sqrt{k}\, \partial \log n_{1} \, e_1 \, .
\end{equation}
For a given $n_{1}$ we can get all other integrals as $n_{1}+c$ with a complex constant $c$. For large $c$ we obtain the infinitesimal transformation
\begin{equation}\label{inftransf}
    \delta (i\partial \f) = - \frac{1}{c}\sqrt{k}\, e^{-\frac{i}{\sqrt{k}}e_1 \f} \,e_1+ \cdots \, .
\end{equation}

In order to generalize this discussion to arbitrary $N$, it is useful to describe the residual gauge symmetries  we just found also using  the parameterization of $a_z$ in terms of a group element $g(z)$ of the form (\ref{puregauge}). From this point of view, their existence may appear somewhat  surprising, since the Gauss decomposition in (\ref{Gaussdec}) is unique for a generic group element $g$ (see e.g.\ \cite{Horn}, Ch.\ 3). We should keep in mind however that $g(z) $ is determined only up to a left multiplication by a constant group element $\tilde \L_0$.  To see if there are residual gauge symmetries, we should therefore verify whether there exist constant group elements $\tilde \L_0$ and  upper triangular group elements $\L_- (z)$ with diagonal entries equal to 1, such that 
\be 
\label{gaugetransf_on_group_elements}
\tilde g = \tilde \L_0 g \L_-
\ee 
is  lower triangular, i.e. 
\be 
(\tilde g)_-=0\,. \label{rescond}
\ee 

Let us once again treat the case $N=2$ in detail. The group elements $h$ and $N_+$ are in this case given by
\be 
h = \left( \begin{array}{cc} e^{i \frac{\e_1}{\sqrt{ k}}\f } &0\\0&  e^{i \frac{\e_2}{\sqrt{ k}}\f }
\end{array}\right)\,, \qquad N_+ =  \left( \begin{array}{cc}1 &0\\n_{1} & 1
\end{array}\right)\label{npNis2}
\ee
where $n_{1}$ is determined by (\ref{Npitoh}) to be a  solution of (\ref{condforF1}). We observe that the equation that determines the component $n_1$ of $N_+$ coincides with the equation~\eqref{condforF1} that determines the finite gauge transformation.
Working out equation (\ref{rescond})  one finds that it is solved by\footnote{This solution is not quite unique in that we are still free to multiply $\tilde \L_0$ on the left by a constant lower triangular matrix, which does not influence the transformation of the field $\pa \f$. Also, the function $n_1$ in $\L_-$  can be $any$ integral of (\ref{condforF1}),  independently of the one chosen for $N_+$.}
\be 
 \tilde \L_0 =  \left( \begin{array}{cc}0 &1\\-{1} & 0
\end{array}\right)\equiv P_{12} \, , \qquad \L_- =  \left( \begin{array}{cc} 1 & \pa (\log n_{1})  \\0&1
\end{array}\right)\, .\label{resNis2}
\ee
We note that $\tilde \L_0$ is essentially a permutation matrix exchanging the first and second rows. 
From inspecting the diagonal elements of $\tilde g$ we find  once again that the field $\pa \f$ transforms under the residual gauge symmetry  as (\ref{largegaugetransfoN2}).

This analysis for $N=2$ can now be straightforwardly generalized to  find the  residual gauge transformations  for arbitrary $N$, where one finds that they are determined by the solutions of the differential equations (\ref{Npitoh}). For example, taking $\tilde \L_0$ to be the permutation matrix $P_{a, a+1}$ exchanging the $a$-th and $(a+1)$-th rows (with an appropriate sign as in (\ref{resNis2})) and $\L_-$ the compensating upper triangular gauge parameter, we 
find the family of residual gauge transformations
\begin{equation}\label{largegaugetransformationN}
P_{a, a+1}: \qquad  i\pa \vec \f \to i\pa \vec \f + \sqrt{k}\, \pa (\log n_a) \vec e_a
\end{equation}
where the $n_a$ solve the equation
\begin{equation}
 \partial n_{a} = -\, e^{-\frac{i}{\sqrt{k}}\, \vec e_a \cdot \vec \f}\, .\label{naint}
\end{equation}
The corresponding infinitesimal transformations are 
\begin{equation}\label{inftransfN}
    \delta_a (i\partial \vec \f) = - \frac{1}{c_a} \sqrt{k}\, e^{-\frac{i}{\sqrt{k}}\,\vec e_a \cdot \vec \f} \,\vec e_a+ \cdots \, .
\end{equation}
More generally, one finds  solutions to (\ref{rescond}) for $\tilde \L_0$ any permutation matrix $P_{ab}, a<b$.
These can however be constructed by composing elementary transformations of the type (\ref{largegaugetransformationN}), since for example $P_{a,a+2} = (P_{a+1,a+2})^{-1} P_{a,a+1} P_{a+1,a+2}$.

The above analysis tells us that, naively, the diagonal gauge allows an $N(N-1)/2$-parameter family of residual proper transformations, depending on the  integration
constants contained in the solution of the system (\ref{Npitoh}). However we have not verified whether, in expressions such as (\ref{largegaugetransformationN}),  the transformed fields are well-defined, i.e.\ sufficiently regular. This is what we are now going to investigate.

\subsection{Gauge orbits}\label{secfinite}

In this section we want  to get a clearer picture of the gauge orbits swept out by the residual gauge transformations in the diagonal gauge. In order to do so, we must first specify which are the field configurations we allow, and then investigate which residual transformations map allowed field configurations into each other.  In what follows we will impose that the $\pa \f^a $ are analytic on the Riemann sphere with only isolated singularities and no branch cuts. The physical motivation behind this assumption is that we want to exclude multi-valued fields, but allow for point-like sources.
  For generic field configurations in this class and generic choices of the parameters entering in the residual gauge transformations such as (\ref{largegaugetransformationN}), the transformed fields  contain branch cuts and are outside of the allowed field space. Therefore our task is to determine for which field configurations and gauge parameters the residual gauge transformations do act within our field space. 
 
 For simplicity of the discussion we will focus on the action of the residual gauge transformations on fields which have singularities only in the origin 
  and infinity, i.e.\ the infinite past and future in radial time. The results of this analysis can 
  be extended straightforwardly to more general configurations of singularities.
We will distinguish between two classes of field configurations, for which the analysis  of the allowed residual transformations  is  qualitatively different.

\subsubsection*{Generic fields: discrete orbits}
Let us consider a generic field configuration $\pa \vec \f$ with singularities only in the origin and at infinity. The exponentials $e^{-\frac{i}{\sqrt{k}}\vec e_a \cdot \vec \f}$ typically have a branch cut running from  $z=0$ to  infinity, and we  see from  \eqref{inftransfN} that  the corresponding infinitesimal transformation is obstructed. Nevertheless  
it is still possible that the finite transformation (\ref{largegaugetransformationN}) is well-defined for a specific choice of $n_a$. 
We therefore ask whether there is any $n_a$ satisfying~\eqref{condforF1} such that $\partial\log n_a$ is again analytic on $\mathbb{C}$ without branch cuts and only with isolated singularities. It turns out that there is precisely one  such solution, given by
\begin{equation}
    n_a = d_a\, \int_{z}^{e^{2\pi i}z} e^{-\frac{i}{\sqrt{k}}\,\vec e_a \cdot  \vec \f (\tilde{z})}d\tilde{z} \, .\label{nintegral}
\end{equation}
Here, the constant $d_a$ is chosen such that $n_a$ solves~\eqref{condforF1}, but it drops out of the transformation~\eqref{largegaugetransfoN2}. This means that  for each fixed $a$, the residual gauge symmetry  (\ref{largegaugetransformationN}) is just a discrete $\mathbb{Z}_2$ symmetry. When these are combined they  generate the symmetric group and relate $N!$ solutions. 

It is instructive, and will be useful for what follows,  to work out how the symmetric group  acts on the simple subclass of `zero mode'  solutions  where the $1$-form $\pa \vec \f$ has  simple poles in the origin and at infinity:  
\be 
i\pa \vec \f = \frac{\vec \a_0}{z}\,.\label{purezero}
\ee
One finds from evaluating  (\ref{nintegral}) that the $a$-th elementary residual symmetry (\ref{largegaugetransformationN}) acts on the zero modes as
\be
\vec \a_0 \to s_a ( \vec \a_0 - \sqrt{k}\, \vec \r ) + \sqrt{k}\, \vec \r \, , \label{a0equiv}
\ee 
where $\vec \r = \sum_a \vec \o_a$ is the Weyl vector and $s_a$ is the $a$-th elementary Weyl reflection defined as 
\be 
s_a (\vec \b) = \vec \b - (\vec e_a \cdot \vec \b)\, \vec e_a\,.\label{elweyl}
\ee 
Therefore, the symmetric group acts as the   Weyl group on the shifted zero mode
$\vec \a_0^{cyl} \equiv \vec \a_0 - \sqrt{k}\, \vec \r$. Note that this quantity can be interpreted as the zero mode after making the conformal transformation to the cylinder (see section \ref{seccondef}), the shift arising from the background charge.

\subsubsection*{Smooth gauge fields: continuous orbits}
In the subspace of field configurations where the exponentials
  $e^{-\frac{i}{\sqrt{k}}\vec e_a \cdot \vec \f}$ have no branch cuts, there is no obstruction for the infinitesimal transformation \eqref{inftransf}, and the residual gauge transformations have the chance of being realized as a continuous symmetry.  Note that this subspace is characterized by a quantization condition on the residue in the origin: 
  \be
{\rm Res}_{z \to 0} (i \pa \vec \f)  \in  \sqrt{k }\, W \, ,\label{a0quantcl}
\ee
where $W$ is the weight lattice.

The condition (\ref{a0quantcl}) can be rephrased in a more insightful way as requiring that the diagonal group element $h(z)$ in the Gauss decomposition (\ref{puregauge}) is single-valued  in the gauge group $SL(N,\CC)/\ZZ_N$. Indeed, from (\ref{gdiag}) we find that single-valuedness of $h(z)$ as $z$ encircles the origin imposes the following condition on $\vec \f$:
\be
\vec \e_i \cdot \vec \f (e^{2 \p i}z ) = \vec \e_i \cdot \vec \f (z )+ 2\p \sqrt{k} \left(
m_i - \frac{m}{N}\right) ,
\ee
where $m_i\in \ZZ$ and $\sum_i m_i = m$. We allowed for $h(z)$ to come back to itself up to a phase $e^{2\p i m / N},\ m \in \ZZ$, since we mod out by the $\ZZ_N$ center. Multiplying by $\vec \e_i$ and summing over $i$
gives, using (\ref{epsprops}),
\be
\vec \f (e^{2 \p i} z) = \vec \f (z )+ 2\p \sqrt{k} \sum_i
m_i \vec \e_i\,.\label{phiper}
\ee
Using the second relation in (\ref{epsprops2}), we see that the lattice spanned by vectors of the form $\sum_i  m_i \vec \e_i$ is nothing but the weight lattice. Therefore single-valuedness of $h(z)$ in the origin is equivalent to (\ref{a0quantcl}). 

Though we saw that the condition (\ref{a0quantcl}) ensures that the infinitesimal residual transformations \eqref{inftransf} are well-defined, it does not guarantee that this remains true for finite residual transformations
such as (\ref{largegaugetransformationN}), since branch cuts may appear at higher order. From inspecting  (\ref{largegaugetransformationN}) one easily sees that this leads to further conditions on the fields.
Indeed, if $e^{-\frac{i}{\sqrt{k}}\vec e_a \cdot \vec \f}$ has a non-trivial residue in $0$, then every solution $n_a$, as well as $\partial \log n_a$, will have a cut and the elementary  residual transformation (\ref{largegaugetransformationN}) is not well-defined. Requiring that also the more general  residual transformations, which can be obtained by composing the elementary ones, are well-defined 
leads to the additional condition that there are no residues in the exponentials of the transformed fields, such as
\be 
\big( n_b (\vec \f)\big)^{- C_{ab}} \,e^{-  \frac{i}{\sqrt{k}}\, \vec e_a \cdot \vec \f}\,,\label{extracond}
 \ee
 where $C_{ab}$ is the Cartan matrix, and so on.
Though we will not attempt to give a complete characterization of the resulting restrictions on field space in this work, it is not hard to see that the full set of extra conditions can be summarized as the
 requirement that the group element $N_+$ in
the Gauss decomposition (\ref{puregauge}) is single-valued. This follows from our observation in the previous subsection  that the differential equations which determine the residual gauge parameters are the same as those determining the components of $N_+$. Indeed from the discussion around~\eqref{gaugetransf_on_group_elements} it can be readily seen that for a single-valued $N_+$ and a given $\tilde{\L}_0$, also $\L_-$ is single-valued.

In summary, we saw that the residual gauge transformations act as  continuous symmetries  on field configurations for which both $h$ and $N_+$, and hence the group element $g$ in (\ref{puregauge}), is single-valued. 
This condition can be interpreted more physically as a smoothness condition for the gauge field $a_z$,
namely that the holonomy of $a_z$ around the origin is trivial in the gauge group $SL(N,\CC)/\ZZ_N$.
Indeed, $a_z$ has trivial holonomy around the origin if
\be
\calp \exp \oint_{S_z} A = g^{-1}(e^{2 \p i} z ) g(z) = e^{2\p i m / N} \cdot \unity  \label{trivhol}
\ee
for some integer $m$, where $S_z$ is a circle centered at the origin and  going through $z$.  Trivial holonomy is therefore equivalent to having $g(z) $ single valued    in
$SL(N,\CC)/\ZZ_N$.

To summarize, we found that the orbits under residual gauge symmetries are rather different depending on whether 
\begin{itemize}
    \item the field configuration $\pa \vec \f$  leads to a singular gauge potential $a_z$, as  is generically the case. The residual gauge transformations then act  discretely, instructing us to identify discrete points in field space. At  such generic points  the
diagonal gauge suffers from a Gribov-type ambiguity \cite{Gribov:1977wm}: the gauge-fixing slice intersects the gauge orbit of the generic field configuration $N!$ times. 
\item the field configuration $\pa \vec \f$ leads to a smooth gauge potential $a_z$, in the sense that the holonomy is trivial.
The residual gauge symmetries act as continuous symmetries, instructing us to identify points on continuous orbits in field space. In this case  a finite-parameter family of  residual gauge symmetries generates a motion tangent to the diagonal gauge slice specified by (\ref{diaggauge}).
\end{itemize} 
In both situations we should quotient our field space by the action of the residual symmetries, and this will be the subject of the next section.
We conclude this section with some further comments:
\begin{itemize}
    \item It is instructive to note that the continuous residual gauge transformations typically introduce additional pole terms in the fields $\pa \vec \f$. For example, starting from the configuration with a pole in the origin and satisfying (\ref{a0quantcl}),
    $i\pa \vec \f = - \sqrt{k }\vec \L /z$ with $\vec \L \in W$, we find that the residual transformations 
    (\ref{largegaugetransformationN}) act as
    \be
P_{a, a+1}: \qquad     i\pa \vec \f  \to i\pa \vec {\tilde \f} =   - \frac{\sqrt{k}\,\vec\L}{z} 
+ \frac{\sqrt{k}\,(\L_a + 1)\, z^{\L_a}  \vec e_a}{z^{\L_a+1} - z_0^{\L_a+1 }}
    \ee
    with $z_0$ an integration constant. The transformed field has additional poles; for example there is now also  a pole in $z_0$ with  residue  
    $\sqrt{k} \,\vec e_a$, which satisfies (\ref{a0quantcl}). 
    
    Only when we let  $z_0 \to 0$, all the  poles merge 
    to a first order pole in the origin, with a residue which is transformed according to  (\ref{a0equiv}). On the weight vector $\vec \L$, the transformation acts as a shifted Weyl reflection
    \be P_{a, a+1}: \qquad    \vec \L \to s_a \cdot \vec \L  \equiv s_a (\vec \L + \vec \r) - \vec \r \,.
    \label{shiftweyl}\ee
    with $s_a$ the $a$-th elementary Weyl reflection, see (\ref{elweyl}). 
    \item One way\footnote{Note that this way of obtaining the  quantization condition (\ref{a0quantcl}) from winding mode quantization for a periodic scalar is by no means unique. For example, in Appendix \ref{appBfield} we discuss an alternate realization, where the periods lie in the rescaled root lattice in the presence of a constant $B$-field.}
to think about the quantization condition (\ref{a0quantcl}) is to view the scalar fields as being periodic  with identifications
\be
\vec \f \sim \vec \f + 2 \p \sqrt{k}\, \vec \L \,, \label{phiperiod}
\ee
with any vector $\vec\L\in W$ in the weight lattice. The value of the residue in (\ref{a0quantcl}) then labels the different winding sectors.
\item Our analysis is reminiscent of the classification of the
coadjoint orbits of the Virasoro group \cite{Witten:1987ty}, where one distinguishes between generic orbits and exceptional ones by the presence of continuous symmetries. 
We will indeed see that the trivial holonomy condition on $a_z$ defines a generalization of the
exceptional orbits to the case of $W_N$ symmetry.\footnote{ See \cite{Bajnok:2000nb} and references therein for a discussion of $W_N$ coadjoint orbits.} In the quantum theory, these will turn out to  correspond to degenerate representations, while the
generic case leads to nondegenerate ones. 
\end{itemize}

\section{Asymptotic symmetries and quantization}

In the previous section we have seen that it is possible to describe asymptotically AdS configurations in the diagonal gauge, but that the latter entails residual gauge symmetries.
To obtain a theory equivalent to one formulated in a Drinfeld-Sokolov gauge, one must therefore quotient out these residual symmetries. This naturally leads to a classical version of the Coulomb gas formalism for CFT's with $W_N$ symmetry developed in \cite{Lukyanov:1990tf}: the generators of the residual gauge symmetries are precisely the screening charges of the Coulomb gas formalism. In order to illustrate this picture, in this section we proceed as follows: we first build a Poisson bracket on the space of boundary excitations, as described by the fields that appear on the diagonal of the gauge connection \eqref{diaggauge}. In sect.~\ref{sec:global} we show that, under reasonable assumptions, one obtains the Poisson bracket of free fields. In sect.~\ref{secresidual} we then build the canonical generator of the residual gauge symmetries on this phase space, obtaining a classical analogue of the screening charges of the Coulomb gas formalism. To support our findings, in sect.~\ref{secWN} we also show that the transformations generated by the screening charges are symmetries of the Miura transform, which relates the diagonal gauge to a specific Drinfeld-Sokolov gauge. This means that the gauge invariant observables on the free field phase space are precisely the $W_N$ charges that one obtains in Drinfeld-Sokolov gauges, thus showing the equivalence of the two approaches. In sect.~\ref{secquant} we eventually show how the subtle classical analysis of the diagonal gauge pays off when moving to quantization: the resulting boundary phase space can indeed be quantized with standard and efficient techniques.

\subsection{Global symmetries and Poisson brackets of free fields}\label{sec:global}
 
To identify the Poisson brackets of the basic fields we now examine the variations \eqref{gaugetransf} of the connection which preserve the diagonal gauge \eqref{diaggauge}. We aim to distinguish between global symmetries ---~which determine the structure of the boundary phase space as we have reviewed in section~\ref{sec:review}~--- and proper gauge symmetries.

We recall that we can write $a_z$ as
\begin{equation} \label{h-decomposition}
a_z = J_{1} + h^{-1}\pa_z h
\end{equation}
with $h$ defined in~\eqref{gdiag}. If one rewrites the gauge parameter as
\begin{equation}
\lambda  = h^{-1} \, \tilde{\lambda}\, h \,,\label{litolt}
\end{equation}
the transformation of $a_z$ then reads
\begin{equation}
\delta_\l a_z =\pa_z \lambda  + [a_z,\lambda] = h^{-1}\left( \pa_z \tilde{\lambda} + [\,h\,J_{1}\,h^{-1},\tilde{\lambda}\,]\right) h\, .\label{deltaresid}
\end{equation}
The diagonal gauge condition~\eqref{diaggauge} is therefore preserved  if and only if the factor in brackets vanishes away from the main diagonal.
In terms of the components $\tilde{\l}_{j,k}$ of the matrix $\tilde{\l}$, this leads to the conditions
\be
\pa_z \tilde \l_{j,k} = e^{ - {\frac{i}{\sqrt{k}}\, \vec e_{j-1} \cdot \vec \f} }\, \tilde \l_{j-1,k} - e^{ - {\frac{i}{\sqrt{k}}\, \vec e_{k} \cdot \vec \f} }\, \tilde \l_{j,k+1} \qquad {\rm for\ } j\neq k\,.\label{eqlt}
\ee
These equations determine  the $ \tilde{\l}_{j,k}$ for $j \neq k$ in terms of the elements on higher diagonals, while the elements on the main diagonal are arbitrary. Eqs.~\eqref{eqlt} therefore do not mix the lower-triangular and strictly upper-triangular parts of $\tilde \l$, and we can treat these cases separately. Actually, these two classes of transformations have
very different interpretations: as we shall see shortly, lower-triangular gauge parameters encode the global symmetries of the system, mapping into each other physically inequivalent solutions, while strictly  upper-triangular $\tilde  \l$ (and hence $ \l$)  generate the proper gauge symmetries discussed in sect.~\ref{sec:diagonalgauge}. 

The lower-triangular gauge parameters preserving the diagonal gauge are of the form
\be \label{lambda}
\l =  {\frac{1}{\sqrt{k}}\sum_{j=1}^N  \vec{\e}_j \cdot (i\, \vec{\xi}(z)) e_{jj} }+ {\rm strictly \ lower\ triangular} \, ,
\ee
 where $\vec \xi (z)$ is a vector of arbitrary holomorphic functions and the remaining strictly lower-triangular part of $\l$
  is determined  by  $\vec \xi$ through the differential equations (\ref{eqlt}). This portion of the resulting gauge parameter depends on the fields: nevertheless, one can still integrate the variation of the charges in \eqref{varQ} because its scalar product with the connection vanishes. This means that, even if the gauge parameter is field dependent, one can still use \eqref{consQ2} to evaluate the asymptotic charges, in complete analogy with what one usually does in the analysis of the global symmetries of Drinfeld-Sokolov gauges (see e.g.~\cite{Banados:1998gg,Campoleoni:2010zq}). The choice of a specific solution of the system of equations \eqref{eqlt}, i.e.\ of the integration constants,  is also immaterial, since they do not affect the connection and they drop out of the charges.
 
Substituting \eqref{lambda} into \eqref{consQ2} taking \eqref{epsprops} into account, the asymptotic charges eventually read 
 \be
Q_{\vec \xi, \vec {\bar \xi}} =\frac{1}{2\p} \oint \left(   \vec \xi \cdot \pa_z \vec  \f \,dz + \vec {\bar \xi} \cdot  \pa_{\bar z} \vec {\bar \f} \,d\bar z\right) , \label{KMcharges}
\ee
 while gauge transformations generated by parameters of the form \eqref{lambda} induce shifts of the fields that only depend on the entries of the main diagonal:
\be
\d_{\vec \xi, \vec {\bar \xi}}\,\pa_z \vec \f =\pa_{ z} \vec \xi\, , \qquad 
\d_{\vec \xi, \vec {\bar \xi}}\,\pa_{\bar z} \vec {\bar \f} = \pa_{\bar z} \vec {\bar \xi}\, . \label{Psivar}
\ee
We now assume that these transformations are generated by the charges as
\be
\d_{\vec \xi, \vec {\bar \xi}} \,\partial_z \vec{\phi} = \{ Q_{\vec \xi , \vec {\bar \xi}} \,, \partial_z \vec{\phi} \} \, , \label{DBvar}
\ee
for a proper Poisson bracket defined on the space of fields in the diagonal gauge.\footnote{If all constraints were second class, this would be the standard Dirac bracket \cite{Brown:1986ed}. In our case, we propose to still apply~\eqref{DBvar} and to then complement the results it gives in such a way to obtain a well defined phase space. This procedure is also supported by the observation that, as discussed in the $sl(2)$ example in section~\ref{secadsas}, the constraint matrix is invertible almost everywhere, and this suggests that the induced Dirac bracket can be extended to the whole space of fields in the diagonal gauge. This reasoning similarly applies to all $sl(N)$ cases: as shown in (3.96) of \cite{Balog:1990mu}, for $sl(N)$ the constraint matrix indeed has a block-diagonal form as in the $sl(2)$ case. The steps needed to obtain the bracket in (3.99) of \cite{Balog:1990mu} are however not as direct as claimed in the latter paper; the non-vanishing block of the constraint matrix is not invertible on the whole constraint surface due to its dependence on phase-space variables.}

We then expand $\vec \f$ and $\vec {\bar \f}$ in Laurent modes
\be
i \pa_z  \f^a = \sum_{m \in \ZZ} \frac{\a^a_m}{z^{m+1}}\, , \qquad 
- i \pa_{\bar z} \bar \f^a = \sum_{m \in \ZZ} \frac{\bar \a^a_m}{\bar z^{m+1}}\, .\label{modeexp}
\ee
Using (\ref{DBvar}) for gauge parameters of the form $\vec{\xi}(z) = \vec{\xi}_n\, z^n$ we find for the modes
\be
 - i \{ \a_{m}^a ,  \a_{n}^b \} = m\, \d^{ab}\d_{m,- n}\,,\qquad 
 -\, i \{\bar  \a_{m}^a , \bar \a_{n}^b \} =  m\, \d^{ab}\d_{m,- n}\,,\qquad 
 \{ \a_{m}^a , \bar \a_{n}^b \} = 0\, .\label{freePB}
\ee
Note that the zero modes $\vec \a_0$ and $\vec{\bar \a}_0$ are `central', in the sense that they Poisson-commute
with everything else. In particular, as it stands there are no canonically conjugate variables to $\vec \a_0, \vec{\bar \a}_0$ in the space of modes. With hindsight, this is not surprising: one cannot expect this procedure to reproduce a bona fide phase space, due to the presence of the residual gauge symmetries discussed in section \ref{sec:residualgaugesymmetries}.

To proceed we then propose to slightly extend the space \eqref{freePB} by introducing zero modes canonically conjugate to $\vec \a_0$ and $\vec{\bar \a}_0$ by hand. These can be introduced  rather naturally  by promoting the constant mode of $\vec \f$ (which does not appear in the gauge connection $a_z$) to a dynamical variable, and to declare that it is shifted by acting  with a constant gauge parameter $\vec \xi$. In other words, we replace (\ref{Psivar}) with its integrated version
\be
\d_{\vec \xi, \vec {\bar \xi}}\,\vec \f = \vec \xi \, , \qquad \d_{\vec \xi, \vec {\bar \xi}} \,\vec {\bar \f} =  \vec {\bar \xi} \, ,\label{Psivar2}
\ee
so that the extra zero-modes can be considered as a sort of Stueckelberg fields.
From the Laurent expansion of $\vec \f$,
\be
i \vec \f = i \vec \f_0 + \vec \a_0 \log z- \sum_{m \in \ZZ_0} \frac{\vec{\a}_m}{m\, z^{m}}\, ,\label{modeexp2}
\ee
and following the procedure above we find, in addition to (\ref{freePB}), the zero-mode Poisson brackets 
\be
\{   \a_{0}^a , \f_{0}^b \} = \d^{ab}\, .\label{freePBzerom}
\ee

At this stage we have a concrete proposal for a boundary phase space for any $sl(N,\mathbb{C})$ Chern-Simons theory satisfying the AdS boundary conditions \eqref{asads} supplemented by the partial gauge fixing of Drinfeld-Sokolov symmetries leading to the diagonal gauge. We know from section~\ref{sec:diagonalgauge} that residual gauge symmetries do exist. The goal of the next subsection is to identify the canonical generators of these symmetries on the phase space defined by \eqref{freePB} and \eqref{freePBzerom}. This will allow us to then identify the observables of the theory with the quantities which Poisson commute with these generators, and to verify that they correspond to the charges computed in Drinfeld-Sokolov gauges.

\subsection{Screening charges as generators of residual gauge symmetries}\label{secresidual}

In the preceding subsection we have formulated Poisson brackets on the space of gauge connections satisfying the diagonal gauge, extended by zero modes $\f_0^a$. In this section we will find canonical generators on this phase space which generate the residual gauge symmetries discussed in section \ref{sec:diagonalgauge}.
Note that in this context the charge formula \eqref{consQ2} is not expected to be of any help in selecting these generators. For instance the variation $\d Q \sim \int \tr \l \d a =0$ vanishes for strictly upper triangular $\l$ and diagonal $\d a$, so that the residual symmetries should be associated to constant charges, compatibly with their interpretation as proper gauge symmetries. We will therefore construct directly the functions on the phase space \eqref{freePB}, \eqref{freePBzerom} that generate via \eqref{DBvar} strictly upper triangular gauge transformations preserving the diagonal gauge.

We have discussed in section~\ref{sec:residualgaugesymmetries} that we can express the residual gauge transformations of the connection $a_z=g^{-1}\partial_z g$ as transformations of the type~\eqref{gaugetransf_on_group_elements} on $g(z)$. Infinitesimal transformations can be described by considering $\tilde{\L}_0 = \unity +\e \tilde{\l}_0$ with a constant strictly upper triangular matrix $\tilde{\l}_0$ and an infinitesimal parameter $\e$, together with $\L_-(z)=\unity + \e \l_-(z)$ such that
\begin{equation}
    \tilde{g}(z) = (\unity +\e \tilde{\l}_0) g(z) (\unity + \e \l_-(z)) = g(z) +\e\, (\tilde{\l}_0 g(z) + g(z) \l_-(z) ) + \dots
\end{equation}
is lower triangular. One can observe that $\l_-(z)$ is completely determined in terms of $g$ and $\tilde{\l}_0$. The infinitesimal action on the gauge connection is then
\begin{equation}
    \d a_z = \partial_z \l_- + [a_z,\l_-] \, .
\end{equation}
A basis of the infinitesimal transformations is given by the generators
\begin{equation}
    \tilde{\l}_0^{(ij)} = e_{ab}  \qquad (a<b)\, .
\end{equation}
Obviously, the Lie algebra that one obtains by considering commutators of such transformations is that of upper triangular matrices, where every element can be generated from the elementary transformations $\tilde{\l}_0^{(a,a+1)}$. For these elementary transformations one can solve for $\l_-$, and one obtains 
\begin{equation}
    \l_-^{(a,a+1)} (z) = - h^{-1}(z)\, e_{a,a+1}\, h(z) = -e^{-\frac{i}{\sqrt{k}}(\vec{\e}_a-\vec{\e}_{a+1})\cdot \vec{\f}(z)}\,e_{a,a+1} \, .
\end{equation}
Denoting by $\d_a$ the corresponding transformation generated by $\l_-^{(a,a+1)}$ we find 
\begin{equation}
    \d_a \big(i\partial_z \vec{\f}\big) = -   \sqrt{k}\, e^{- \frac{i}{\sqrt{k}}\, \vec e_a\cdot \vec \f} \,\vec e_a\,, \qquad  a = 1, \ldots , N-1 \label{extrasymms1}\, .
\end{equation}
When we look for charges $Q_{ab}$ that generate the transformations associated to $\tilde{\l}_0^{ab}$, it is enough to find charges $S_a=Q_{a,a+1}$ that generate the elementary transformations~\eqref{extrasymms1}, because all other charges $Q_{ab}$ can then be obtained by forming Poisson brackets of the elementary charges. Such charges are given by
\be 
S_{a} =  \frac{k}{2 \p} \oint dz  \, e^{- \frac{i}{\sqrt{k}}\, \vec e_a\cdot \vec \f(z)} \, .
\label{classscreening}
\ee
Note that $\vec{\f}$ contains a logarithm, and therefore we have to specify the contour and the branch of the logarithm. As the logarithm only appears together with the zero mode $\a_0^a$, different choices here will only lead to a modified transformation of the zero mode $\f_0^a$. Let us now check that the charges generate the desired transformation,
\be
\{S_{a} , i \pa_z \vec \f \} = -\sqrt{k} \, e^{- \frac{i}{\sqrt{k}}\, \vec e_a\cdot \vec \f} \,\vec e_a\, .
\ee
We first observe that
\begin{equation}
    \big\{ e^{-\frac{i}{\sqrt{k}}\,\vec{e}_a \cdot \vec{\f}(z)} , \vec{\a}_m  \big\} = \frac{i}{\sqrt{k}}\, \vec{e}_a\, z^m \, e^{-\frac{i}{\sqrt{k}}\,\vec{e}_a \cdot \vec{\f}(z)} \, .
\end{equation}
Then
\begin{align}
    \big\{ S_a , i\partial_z \vec{\f}(z) \big\} & = \sum_{m\in\mathbb{Z}} z^{-m-1} \big\{ S_a , \vec{\a}_m \big\}\\
    &= -\frac{\sqrt{k}}{2\pi i} \,\vec{e}_a \sum_{m\in\mathbb{Z}} z^{-m-1} \oint_{|z'|\,=\,\text{const.}} \!\!\!\!\!\!\!\!\!dz'\, z'^{m}\, e^{-\frac{i}{\sqrt{k}}\,\vec{e}_a \cdot \vec{\f}(z')}  \\
    &= -\frac{\sqrt{k}}{2\pi i} \,\vec{e}_a \Bigg( \oint_{|z'|>|z|} dz'\, \sum_{m<0} z^{-m-1}z'^{m}\,  e^{-\frac{i}{\sqrt{k}}\,\vec{e}_a \cdot \vec{\f}(z')}  \nonumber \\
    & \qquad \qquad \qquad +\oint_{|z'|<|z|} dz'\, \sum_{m\geq 0 } z^{-m-1}z'^{m}\,  e^{-\frac{i}{\sqrt{k}}\,\vec{e}_a \cdot \vec{\f}(z')} \Bigg)\\
    &= -\frac{\sqrt{k}}{2\pi i} \,\vec{e}_a \bigg( \oint_{|z'|>|z|} - \oint_{|z'|<|z|} \bigg) dz'\, \frac{1}{z'-z}\, e^{-\frac{i}{\sqrt{k}}\,\vec{e}_a \cdot \vec{\f}(z')} \\[6pt]
    &= -\sqrt{k}\, e^{-\frac{i}{\sqrt{k}}\,\vec{e}_a \cdot \vec{\f}(z)}\, \vec{e}_a \, .
\end{align} 
Note that we have assumed here that $e^{-\frac{i}{\sqrt{k}}\vec{e}_a \cdot \vec{\f}(z')}$ does not have a cut. As we have seen in section~\ref{secfinite} this is precisely satisfied for those configurations where the infinitesimal residual gauge transformation is unobstructed.

By a similar computation, one can also determine the transformation of $\vec{\f}$, including the zero mode, which will depend on the choice of contour in~\eqref{classscreening}. This reflects the freedom that we have when extending the transformations from $a_z$, labelled by $\partial_z \vec{\f}$, to $\vec{\f}$, and the precise relation between the transformation of $\vec{\f}$ and the choice of contour will not be important in the following.

We conclude that we have indeed found the canonical generators of the residual gauge transformations. Note that they vanish on configurations for which the gauge transformations are unobstructed: we observed in section~\ref{secfinite} that continuous gauge orbits only exist when $e^{-\frac{i}{\sqrt{k}}\vec e_a\cdot \vec\f}$ has a vanishing residue. Restricting the field space to such configurations can therefore be thought of as implementing the classical screening charges as first-class constraints. 

For the physical symmetries of our theory this implies that out of the global symmetries that we found in the previous subsection only those which Poisson commute with the charges $S_a$ should be considered as true physical symmetries. As we will see in the following subsection, these symmetries form a classical $W_N$ algebra. We will refer to the charges $S_a$ as screening charges since, in a sense to be made precise in section \ref{secquant} below, they are  a classical limit
of  the  screening charges of the Coulomb gas formalism, see \cite{Fateev:1987zh}.

\subsection{Classical $W_N$ algebra}\label{secWN}

Having discussed the presence of residual symmetries in the diagonal gauge and having  derived the screening charges (\ref{classscreening}) which generate them, we now turn to  the construction of gauge-invariant observables which Poisson-commute with the screening charges. In the class of  Drinfeld-Sokolov gauges which we reviewed in section \ref{secadsas},  the DS gauge freedom is completely fixed, and therefore the   variables which  parameterize the reduced phase space  in these gauges should automatically provide us with gauge-invariant observables. It is well-known that these variables are the modes of the $W_N$ currents which form a classical  $W_N$ algebra under Poisson brackets. As a consistency check, we will now verify explicitly that these indeed Poisson-commute with the screening charges (\ref{classscreening}). As an added bonus this analysis will tell us how to relate the diagonal gauge to the DS gauges. 

We will focus here on a particularly convenient DS gauge, which is sometimes called the `U-gauge'. In this case the gauge field takes the form
\be
a_z^U= J_1 - \sum_{j=2}^N (-\sqrt{k})^{-j} U^j (z) e_{1,j}\, .\label{Ugauge}
\ee
It can be shown (see e.g.\ \cite{Bershadsky:1989mf}) that the relation between the $U_j$ and the  diagonal gauge fields $\f^a$  is given by the classical Miura transformation
\be
M\equiv \left( \pa + \frac{\vec \e_1 \cdot ( i \pa \vec \f)}{\sqrt{ k}}\right) \cdots \left( \pa + \frac{\vec \e_N \cdot ( i \pa \vec \f)}{\sqrt{ k}}\right) = \pa^N - \sum_{j=2}^N (-\sqrt{k})^{-j} U_j \pa^{N-j}\, .\label{classmiura}
\ee
By comparing the coefficients of $\pa^{N-j}$ in these two ways of writing the differential operator $M$, we obtain expressions for the $W_N$ currents $U^i$ in terms of the diagonal gauge fields $\pa \vec \f$.
For example, for $U^2$ and $U^3$ one finds, denoting for brevity $\psi'_i \equiv\vec \e_i \cdot \pa \vec \f$,
\bea
U^2 &=& \sum_{i<j}\psi_i' \psi_j' -i\sqrt{k}\sum_j (j-1) \psi_j''\nonu
&=& - \half \vec \f' \cdot \vec \f' + i \sqrt{k} \vec \r \cdot \vec \f''\label{U2class}\\
U^3 &=&-i \sum_{i<j<k} \psi_i' \psi_j' \psi_k'  - \sqrt{k} \sum_{i<j} \left((i-1) \pa_z (\psi_i' \psi_j')+ (j-i-1) \psi_i' \psi_j''\right)\nonumber\\
&& + \frac{i k}{2}\sum_j (j-1) (j-2) \psi_j'''\, .
\eea
In the second line we have used
\begin{align}
\sum_{i<j} \e_i^a \e_j^b &= - \half \d^{ab}\,, &
\sum_j j \vec \e_j  =&- \vec \r
\end{align}
where $\vec \r$ is the Weyl vector.
 We note that the quantity $U^2$ has the form of a free field  stress tensor in the presence of a background charge.

As explained at the beginning of this section,  the $W_N$-currents $U^i$ should be completely  invariant under DS gauge transformations by construction, which means in particular that they should have vanishing Poisson brackets with the classical screening charges (\ref{classscreening}):
\be
\{ S_a , U^i (z) \} = \d_a U^i (z) =0\, ,\label{Poisson0}
\ee
 with $\d_a$ defined in (\ref{extrasymms1}).
To check this, we show that the variation $\d_a$ of the left-hand side of (\ref{classmiura}) vanishes. Let us illustrate this for $\d_1$: 
 \begin{align}
 \d_1 (LHS) &= \left[-e^{- \frac{i}{\sqrt{k}} \vec e_1\cdot \vec \f} \left( \pa + \frac{i\vec \e_2\cdot \partial\vec\f}{\sqrt{ k}}\right) +  \left( \pa + \frac{i\vec \e_1\cdot \partial\vec\f}{\sqrt{ k}}\right)e^{- \frac{i}{\sqrt{k}} \vec e_1\cdot \vec \f} \right]\nonumber \\ 
 & \quad \times \left( \pa + \frac{i\vec \e_3\cdot \partial \vec \f}{\sqrt{ k}}\right) \cdots \left( \pa + \frac{i\vec \e_N\cdot \partial\vec \f}{\sqrt{ k}}\right)
\end{align}
and one checks using (\ref{epsprops2}) that the operator in square brackets vanishes.
This classical argument mirrors a similar proof for  the quantum screening charges in \cite{Fateev:1987zh}. Conversely, we demonstrate in appendix~\ref{appMiura} that the screening transformations constitute the most general infinitesimal symmetries of the left hand side of the  Miura transformation~(\ref{classmiura}).

It is well-known \cite{Balog:1990mu} that the currents $U^i (z)$   form a classical $W_N$ algebra under Poisson brackets, with
$U^2$ playing the role of the classical Virasoro stress tensor.  A special feature of the U-gauge is that the nonlinearities in the right-hand side of the Poisson bracket algebra is at most quadratic in the $U^i$.
From the analysis above we see that this classical $W_N$ algebra arises here as the Poisson-commutant of a set of screening charges (\ref{classscreening}), which is again a classical limit  of the Coulomb gas description of the quantum $W_N$ algebra. As a further check, we compute the classical central charge of this classical $W_N$ algebra.
Expanding $U^2$ in modes,
\be
U^2 = \sum_{m\in \ZZ} \frac{L_m}{z^{m+2}}\, ,
\ee
we obtain
\be
L_m = \half \sum_{n\in \ZZ} \vec \a_n\cdot \vec  \a_{m-n} - \sqrt{k} (m+1) \vec \r \cdot \a_m\, .\label{classT}
\ee
Using (\ref{freePB}) one checks that their Poisson brackets give the classical Virasoro algebra
\be
- i \{ L_m, L_n \} = (m-n)L_{m+n} + \frac{c_{cl}}{12}\, m (m^2-1)  \d_{m+n,0}\label{virclass}
\ee
with classical central charge
\be
c_{cl } = - 12 k\, \vec \r \cdot \vec \r = - N(N^2-1) k\label{ccl}\, .
\ee
From (\ref{kitoGN}) we see that the classical central charge takes the Brown-Henneaux  \cite{Brown:1986nw} value $c_{cl} = \frac{3 l}{2 G_N}$ when expressed in terms of Newton's constant.

\subsection{Quantization in diagonal gauge}\label{secquant}

We now turn to the quantization of the higher-spin gravity theory in the diagonal gauge. At the classical level, we found a set of variables $\f^a$ satisfying free field Poisson brackets. Due to the residual gauge freedom in this gauge, we also found some additional structure:
the screening charges $S_a$ which generate residual  gauge symmetries, and a set of
 gauge-invariant observables $U^i$ which generate a classical $W_N$ algebra.   
Since our goal is to obtain a quantum theory which is equivalent to  the one obtained by quantizing  in a DS gauge, we have to preserve  this additional structure also at the quantum level.

 We start by replacing the Dirac brackets (\ref{freePB}) with operator commutation relations, $- i \{ \ ,  \ \} \to [\ ,\ ]$, leading to 
\bea
[  \hat \a_{m}^a , \hat\a_{n}^b ] &=& m\, \d^{ab}\d_{m,- n}\, , \qquad [\hat \a_0^a, \hat \f^b_0 ] = - i\, \d^{ab}\,.\label{freecomm}
\eea
The associated free field operators, defined through their Laurent expansions as in (\ref{modeexp}), will be denoted by $\hat{\vec  \f}  (z)$.

Next we set out to construct  quantum screening operators $\hat S_a$  and $W_N$ currents $\hat U^i(z)$.   As in (\ref{classmiura}) it will be convenient to package the latter into a quantum Miura
operator $\hat M$ which should satisfy 
\be
[  \hat S_a , \hat M (z)] =0\, ,\label{commcond}
\ee
and such that $\hat M$ and $\hat S_a$ reduce to their classical versions (\ref{classscreening}, \ref{classmiura}) in the classical large $k$ limit.

Starting with the quantum screening operators, a natural choice is to take them to be the normal-ordered operators
\be\hat S_a =\frac{k}{2\p} \oint dz : e^{- \frac{i}{\sqrt{k}}\, \vec e_a\cdot \hat{\vec \f}}:\label{screening+}\, .
\ee
 Note that the coefficient of the exponential  cannot receive $1/k$ corrections if we impose the periodicity (\ref{phiperiod}) on the fields $\vec \f$ which guarantees that  the gauge field has trivial holonomy.

With these screening charges, we learn from the literature on quantum Drinfeld-Sokolov reduction \cite{Fateev:1987zh}\footnote{We follow here the conventions of \cite{Bouwknegt:1992wg}.}
that the Miura operator $\hat M$ is the normal ordered operator version of $M (z)$ in (\ref{classmiura}), while allowing  the constant which has the interpretation of a background charge to receive $1/k$ corrections.\footnote{This fact was already recognized in the early work \cite{Curtright:1982gt} in the context of Liouville theory:  the background charge in the stress tensor has to be shifted in order for the Liouville potential to remain a weight (1,1) primary in the quantum theory.}
Concretely this means that $\hat M$ is of the form
\bea
\hat M(z) &=& :\left(\pa - \frac{\vec \e_1 \cdot (i \pa \hat{\vec \f})}{\tilde\a_0} \right) \cdots \left(\pa - \frac{\vec \e_N \cdot (i \pa \hat{\vec \f})}{\tilde\a_0} \right):\label{qMiura} \\
&=&\pa^N - \sum_{j=2}^N (\tilde\a_0)^{-j}\hat U^j(z)  \pa^{N-j}
\eea
where the second line defines the quantum $W_N$ currents.
The parameter $\tilde \a_0$ denotes the quantum corrected background charge, which from (\ref{classmiura}) must behave  for large $k$ as
\be
\tilde \a_0 = - \sqrt {k}+\calo (1)\, .
\ee
One should keep in mind that with our current field normalization, the classical limit is $k \to - \infty$ while keeping   $\hat{\vec \f}/\sqrt{k}$ fixed, so that the operators $\hat M(z)$ and  $\hat S_a$ have well-defined limits. 

 A quantum version of our classical argument in section \ref{secWN}, see e.g.\  eq.\ (6.46) in  \cite{Bouwknegt:1992wg}, shows that (\ref{commcond}) holds provided we take
\be
\tilde \a_0 = - \sqrt{k} + \frac{1}{\sqrt{k}}\, .\label{backgchargeshift}
\ee
In the literature (see e.g.\  \cite{Bouwknegt:1992wg})  often  constants $\a_+$ and $\a_-$  are introduced which are defined by  $\a_+ \a_- = -1,\ \tilde \a_0 = \a_+ +\a_- $. In our setup these  are related to $k$ as
\be 
\a_- = -\sqrt{k}\,,\qquad  \a_+ = 1/\sqrt{k}\, .\label{apm}
\ee 
It can be shown that the operators $\hat U^j$ form a
quantum $W_N$ algebra at central charge
\be
c = (N-1)\left(1 - N(N+1)\tilde \a_0^2\right) .
\ee

It is important to note that the quantum Miura operator (\ref{qMiura}) possesses the symmetry $\a_+ \leftrightarrow \a_-$, or
\be
\sqrt{k} \leftrightarrow - \frac{1}{\sqrt{k}}\label{Z2symm}
\ee
which is not visible at the classical level. An immediate consequence is that the quantum theory possesses a second set of screening operators which commute with the $W_N$ currents, namely
\be
\hat {\tilde S}_a =\frac{k}{2\p} \oint dz : e^{ {i \sqrt{k}} \,\vec e_a\cdot \vec \f}:\,.\label{screening-}
\ee
We will comment on further ramifications of this quantum symmetry in the discussion in section \ref{secdisc}.

\section{Application: quantization of conical solutions}\label{seccondef}

We argued in  section \ref{secfinite} that a special role is played by the field configurations $\pa \vec \f$  which lead to a gauge field with trivial holonomy, as these possess a continuous family of residual gauge symmetries.
In this section we will study in detail the class of such configurations which have only a simple pole in the origin and at infinity, i.e.\ where only the zero mode in (\ref{modeexp}) is turned on.
 These turn out to be precisely the conical surplus solutions studied in \cite{Castro:2011iw}. We will also discuss the quantization of these solutions which is  straightforward in the current variables. This will allow us to confirm the conjecture of \cite{Castro:2011iw,Perlmutter:2012ds}
that these solutions correspond to a specific subset of degenerate primaries of the $W_N$ algebra. 

\subsection{Winding sectors and conical solutions}

In this section we focus on classical pure zero mode solutions, where $\vec \a_m = 0$ for $m \neq 0$.
Before restricting to the solutions satisfying (\ref{a0quantcl}),   we would like to  comment on the $L_0$-spectrum of generic pure zero mode solutions.  A useful quantity is $L_0 - \frac{c_{cl}}{24}$, which indicates if the solution lies below the black hole threshold  (when it is negative) or above  it (when it is positive).
From (\ref{classT}) we have
\be
L_0 - \frac{c_{cl}}{24} = \half \left(\vec \a_0 - \sqrt{k} \vec \r\right)^2 \equiv \half \big(\vec \a_0^{cyl} \big)^2\, .\label{l0}
\ee
Here, $\vec \a_0^{cyl}= \vec \a_0 - \sqrt{k}\, \vec \r$ can be interpreted as the zero-mode on the cylinder, which receives a shift due to the background charge as we will see in (\ref{tocyl}) below. For real $\vec \a_0^{cyl}$, which was assumed in  \cite{Balog:1990mu}, one obtains solutions above the black hole threshold.   However, the quantization condition 
 (\ref{a0quantcl}) leads to  imaginary  $\vec \a_0$  and $\vec \a_0^{cyl}$ (recall that $\sqrt{k}$ is imaginary in the regime of interest), and therefore the solutions we are interested in here lie below the black hole threshold.

Now, let us specialize to the zero mode solutions where $\vec \a_0$ is quantized according to  (\ref{a0quantcl}), i.e.
\be
\vec \a_0= - \sqrt{k }\, \vec \L \, , \ \vec \a_{m\neq 0} =0\,, \qquad 
i \vec \f =- \sqrt{k}\, \vec \L \log z\, ,\label{windings2}
\ee
where $\vec \L$ is an arbitrary element of the weight lattice, and the arbitrary sign has been introduced for later convenience. As discussed at the end of section \ref{secfinite}, these can be viewed as winding solutions, where $\vec \L$ labels the winding sector.

The analysis in section \ref{secfinite} also showed that certain finite residual gauge transformations relate different values of the weight vector $\vec \L$, which therefore represent the same physical state. 
From (\ref{shiftweyl}), these symmetries  act on   $\vec \L$ 
as  shifted Weyl reflections, 
\be
\vec{\L} \sim w\cdot \vec{\L} = w(\vec{\L} + \vec \r) - \vec \r\, ,\label{shiftedweyl}
\ee
where $w(\L) $ is an arbitrary ordinary Weyl reflection.
This freedom can be used to take   $\vec \L +\vec \r$ to be a dominant weight. In terms of the Dynkin labels,     
\be 
\vec \L =  \sum_{a=1}^{N-1} \L^a \vec \o_a\, ,  \label{dynklab} 
\ee 
this means that we can take
\be 
\L^a \geq -1 \quad \text{for}\ a=1,\dots ,N-1  
\ee
 without loss of generality.

 The analysis in section \ref{secfinite} showed furthermore that there are extra requirements for the gauge field to have trivial holonomy, which can be summarized in the requirement that the matrix $N_+$, which  satisfies (\ref{Npitoh}), is single-valued. This matrix is lower triangular with 1's on the diagonal,
 and working out (\ref{Npitoh}) for the winding solutions (\ref{windings2}) one finds that its components $n_{ab}$ $(a>b)$ should satisfy:
 \bea
 \pa n_{a+1,a} &=& - z^{\L^a}\\
 \pa n_{a + l, a} &=&  - z^{\L^a} n_{a + l, a+ 1} \qquad {\rm for\ } l\geq 2\, .
 \eea
The first line re-states (\ref{naint}).  In order for $N_+$ to be single-valued, the right hand side of these equations should not have a residue in $z=0$. This will be the case if
none of the Dynkin labels  $\L^a$ equals $-1$, and therefore we will further restrict ourselves to the case
where 
\be 
\L^a \geq 0 \qquad \text{for}\ a=1,\dots ,N-1 \, .\label{dynkpos}
\ee

The classical $W_N$ charges of the winding solutions~(\ref{windings2},\ref{dynkpos})   can be computed from (\ref{classmiura}), and in particular for the energy $L_0$  we get, using (\ref{classT}) or (\ref{l0}),
\be
L_0 = k \,\calc_2 (\vec \L )\label{encon}
\ee
where $\calc_2 (\vec \L )= \half \vec \L  \cdot (\vec \L  + 2 \vec \r)$ is the value of the quadratic
Casimir of $sl(N)$. The restriction (\ref{dynkpos}) implies that the  solutions under consideration have energies (\ref{encon}) smaller than or equal to   the energy of the $AdS$ vacuum with $\vec \L =0$.

We arrived at the winding solutions (\ref{windings2}, \ref{dynkpos}) by  requiring the gauge field to have trivial holonomy. We now relate these solutions to the ones studied in \cite{Castro:2011iw}, which  classified the solutions  with trivial holonomy  which can be brought into a Drinfeld-Sokolov gauge by a regular gauge transformation. The outcome was a class of solutions which can be viewed as
generalized conical defects. Since the analysis of  \cite{Castro:2011iw} was performed for  a cylindrical
boundary,  we  expect that our  winding solutions (\ref{windings2},\ref{dynkpos}) 
become precisely the conical solutions of \cite{Castro:2011iw} upon conformal mapping to the cylinder. Let us show in more detail   that this is indeed the case.

As we already observed in  (\ref{U2class}), the  stress tensor for the fields $\vec \f$ contains a background charge term, which implies that $ \pa_z\vec \f $  fields do not transform as primaries under  conformal transformations. Under a finite conformal transformation,
\be z \to w = f(z)\, ,\label{conf}
\ee
the fields transform as\footnote{This can be derived from requiring  invariance of the connection $A = J_0 d\r + \left( e^\r J_1  + \frac{1}{\sqrt{k}}\sum_j \vec \e_j \cdot (i \pa_z \vec \f)\right) dz$. Indeed, one checks that this leads to (\ref{conf}, \ref{conftransfo}) and $\r \to \r - \log f'$.}
\be
i \pa_z \vec \f \to i \pa_w \vec {\tilde \f} = (f')^{-1}  i \pa_z \vec \f + \sqrt{k} { (f')^{-2} f'' } \vec \r\, .\label{conftransfo}
\ee
Applying this to the map from the plane to the cylinder,
\be
w = i \log z\, ,
\ee
we find that a pure zero mode solution maps to
\be
 i \pa_w \vec {\tilde \f} = - i \vec \a_0^{cyl}
\ee
 with 
\be
\vec \a_0^{cyl} = \vec \a_0 - \sqrt{k} \vec \r \label{tocyl} \, .
\ee
In the case of the winding solutions with $\vec \a_0$ given in (\ref{windings2}) one finds $i \pa_w \vec {\tilde \f} =  i \sqrt{k}  (\vec \L + \vec \r)$, and 
 the corresponding gauge connection on the cylinder is
 \be
 \tilde a_w = J_1 + i \sum_j \left(\L^j - \frac{\sum_k \L^k}{N} - j + \frac{N+1}{2} \right)e_{jj} \label{acyl}
 \ee
 where the $\L^j$ are the expansion coefficients  in the $\vec \e_j$ basis: $\vec \L = \sum_{j=1}^N \L^j \vec \e_j$. These are only defined up to an overall  shift, and after appropriately fixing this freedom they can be identified with  the number of boxes in the $j$-th row of the  Young diagram of the
 representation with highest weight $\vec \L$. The connection (\ref{acyl}) is gauge-equivalent to that of the conical solutions constructed in  \cite{Castro:2011iw}, in particular it can be diagonalized and then coincides with  (5.19) in  \cite{Castro:2011iw}.
 The extra condition (\ref{dynkpos}) arose in \cite{Castro:2011iw} from requiring that the solution can be brought to a Drinfeld-Sokolov gauge by a regular gauge transformation.

To conclude this section, we comment on the fact that the winding solutions (\ref{windings2}, \ref{dynkpos}) possess some properties which are suggestive   of  an interpretation as solitons of the theory: their gauge field satisfies a smoothness condition, they have   finite energy which scales with the coupling $k$ as in (\ref{encon}) and are characterized by a topological winding vector $\vec \L$.
One should however not push this analogy too far since they have energies below
the AdS vacuum, and it is possible to show that they also possess unstable directions \cite{Raeymaekers:2014kea}.
In the quantum theory, the latter property will be reflected  in the fact that the corresponding 
quantum states are primaries of $W_N$ representations that are nonunitary in the large central charge limit.

\subsection{Classical symmetries and null vectors}\label{secclassnull}

As we will see below, the winding solutions lead, upon quantization, to degenerate primaries which possess a large number of null descendants.  Before getting to this point, we would like to see first how this property manifests itself already  at the classical level. As was stressed in \cite{Castro:2011ui}, \cite{Perlmutter:2012ds}, null vectors are a quantum manifestation of symmetries of the corresponding classical  solution. For example, the $sl(2,\RR)  \times sl(2,\RR) $ symmetry of the global AdS background is directly related to the fact that the corresponding quantum state, the vacuum,  has null vectors  generated by acting with  $L_{-1}$ and $\bar L_{-1}$.

We will now see that the winding solutions   possess many symmetries.  Indeed, as we showed in section  \ref{secfinite}, the winding  solutions (\ref{windings2}) with (\ref{dynkpos}) allow for a continuous family of residual gauge transformations. These do not change the physical state and therefore generate symmetries of the solution.
To find the corresponding infinitesimal symmetry generators, we evaluate the screening charges (\ref{classscreening})
on a configuration $\vec \f$ in the vicinity of a winding solution (\ref{windings2}) with
$\vec \a_0= - \sqrt{k } \,\vec \L$, and expand the result to linear order in the modes. Doing this we obtain
\bea 
N_a (\vec \L) &=& i k \,e^{- \frac{i}{\sqrt{k}}\, \vec e_a \cdot \vec\f_0} \,{\rm Res}_{z\to 0} \left.\left( z^{\L^a} \exp  \left( \frac{1}{\sqrt{k}}
\sum_{m\neq 0} \frac{\vec e_a \cdot \vec \a_m}{m\, z^m}\right)\right)\right|_{\rm lin} \\
&=& i \sqrt{k} \,\frac{\vec e_a  \cdot \vec \a_{\L^a + 1}}{{\L^a + 1}}\, .\label{posnull}
\eea
We note that
the $N_a (\vec \L)$ obtained in this way have  $L_0$-levels equal to $-\L^a -1$, which is negative due to (\ref{dynkpos}). 
The winding solutions also possess symmetry generators at positive levels.  To obtain these, we use the fact that we can act with the shifted Weyl reflections (\ref{shiftedweyl})  to give a different representative of the same physical solution. In particular, following \cite{Fateev:1987zh}, we make use of the longest element of the Weyl group, $w_0$, which acts on the Dynkin labels as:
\be
(w_0 (\vec \L))^a = - \L^{N-1-a} = -(\L^*)^a \label{w0}
\ee
(where $\vec \L^*$ denotes the weight of the  representation conjugate to $\vec \L$). The corresponding shifted Weyl reflection is
\be
w_0 \cdot \vec \L = - \vec \L^* - 2\vec \rho \, .
\ee
The resulting infinitesimal symmetry generator $N_a (w_0 \cdot \vec \L)$ is 
\be
N_a (w_0 \cdot \vec \L ) =- i \sqrt{k} \,\frac{\vec e_a  \cdot \vec \a_{-\L^{N-1-a} - 1}}{{\L^{N-1-a} + 1}}\, . \label{negnull}
\ee

The present method of obtaining symmetry generators by evaluating the screening charges completely mimics the construction of null vectors in degenerate representations of the quantum $W_N$ algebra, as we discuss in Appendix \ref{appsymm}. In particular our classical construction (\ref{negnull}) gives  $N-1$ basic symmetry generators at  levels  $\L^b +1, b = 1, \ldots, N-1,$ which  are precisely  the levels at which  null vectors appear in certain  degenerate representations. 
The result (\ref{negnull}) can also be derived from an alternative method which was used in
\cite{Perlmutter:2012ds}, where the symmetries of the winding solutions were constructed directly, using the pure gauge form  of $a_z$ in (\ref{puregauge}).

\subsection{Quantization of conical solutions}
Now let us turn to the quantization of the winding solutions (\ref{windings2}).
Here, by quantization of a classical solution $\vec \f$ we mean the identification of a state
in the quantum state space on which the field operator  $\hat{\vec \f}$ has the  eigenvalue  $\vec \f$.
In our case the state space consists of Fock spaces built on vacuum states $|\vec \a_0, 0\rangle$ which satisfy
\be
\hat{\vec \a}_0 |\vec \a_0, 0\rangle = \vec \a_0 |\vec \a_0, 0\rangle\,, \qquad \hat{\vec \a}_m |\vec \a_0, 0\rangle = 0 \ \ {\rm for\ }m>0\,  .
\ee

Let us first review some of the properties of the vacuum states $|\vec \a_0, 0\rangle$  which will be useful below. From (\ref{qMiura}) it follows that they are  $W_N$ primaries, i.e.\  they satisfy $\hat U^s_m |\vec \a_0, 0\rangle =0$ for $m>0$. We will denote the corresponding eigenvalues of the zero modes of the $W_N$-currents as
\be 
\hat U^s_0| \vec \a_0, 0\rangle  \equiv \D_s (\vec \a_0) | \vec \a_0, 0\rangle\, .
\ee
These can be computed by   letting the Miura operator (\ref{qMiura}) act on a vacuum state $| \vec \a_0, 0\rangle $ and inspecting the terms which are singular  as $z\to 0$. This leads to
\be 
\left(\pa - \frac{\vec \e_1 \cdot  \vec \a_0}{\tilde \a_0 z} \right) \cdots \left(\pa - \frac{\vec \e_N \cdot   \vec \a_0}{\tilde\a_0 z} \right)
= \pa^N - \sum_{j=2}^N (\tilde\a_0)^{-j} \D_j ( \vec \a_0) z^{-j} \pa^{N-j}\, .
\ee
This  equation for the $\D_j ( \vec \a_0)$  can be solved to give \cite{Bouwknegt:1992wg}
\be
\D_s ( \vec \a_0) = (-1)^{s-1} \sum_{i_1 < \cdots < i_s} \prod_{j=1}^s \left(  \vec \e_{i_j} \cdot \vec \a_0 + (s-j)\tilde\a_0 \right) . \label{WNcharges}
\ee

Now consider a generic classical 
 field configuration $\vec \f (z)$ which is sufficiently regular for large $z$, in the sense that only the modes $\vec \a_m$
for $m \geq 0$ are turned on. According to the definition above, the   associated quantum state will typically  be a  coherent state.  However, when only the zero mode $\vec \a_0$ is nonvanishing, the corresponding  quantum state is simply the vacuum state $|\vec \a_0, 0\rangle$.
For generic values of $\vec \a_0$, this is a generic $W_N$ primary which does not have any null descendants.
For   the winding solutions (\ref{windings2}, \ref{dynkpos}) however,  the zero mode $\vec \a_0$ takes on the quantized values $\a_- \vec \L$, with $\vec \L$ a dominant weight. They correspond to  the vacuum states
\be
|\a_- \vec \L, 0\rangle\, .\label{windings3}
\ee
The corresponding $W_N$ charges are, from (\ref{WNcharges}),
\be
\D_s (\a_- \vec \L) = (-1)^{s-1} \sum_{i_1 < \cdots < i_s} \prod_{j=1}^s \left( \a_- \vec \e_{i_j} \cdot \vec \L + (s-j)\tilde \a_0 \right) , \label{WNchargescon}
\ee
in particular, the conformal weights read
\be
\D_2 (\a_- \vec \L)= \frac{\a_-}{2}\, \vec \L \cdot \big(\a_-  \vec \L + 2  \tilde\a_0 \vec \r\, \big)\,.
\ee

From these observations we can immediately identify the winding solutions with  the primaries of a subset of completely  degenerate representations of the $W_N$ algebra. Here, the term completely degenerate refers to a representation which contains at least $N-1$ singular vectors (i.e.\ primary null vectors). 
 Completely degenerate $W_N$ representations are labelled by two 
 $sl(N)$ weight
 vectors $(\vec \L', \vec \L)$, and in the free field state space   the corresponding $W_N$ primaries are \cite{Bouwknegt:1992wg}
\be
 |\a_+ \vec \L' +\a_- \vec \L, 0\rangle\, .
 \ee
Therefore the conical/winding   solutions (\ref{windings3}) correspond precisely to the $(0,\vec \L)$  degenerate primaries. The fact that the representations built on these primaries are completely degenerate can be seen from the  explicit construction of the null vectors \cite{Fateev:1987zh}, which we review in Appendix \ref{appsymm}. The operators which create the null vectors reduce  in the classical limit to the symmetry generators we found in (\ref{negnull}).

These considerations give a quantum check, to all orders in an  expansion in  $1/k$, of the conjecture \cite{Castro:2011iw,Perlmutter:2012ds} that   the conical/winding   solutions (\ref{windings2}) correspond to the $(0, \vec \L)$  degenerate primaries.
Previous checks of the conjecture include matching all the $W_N$ charges (\ref{WNchargescon}) in the classical, large $k$ limit \cite{Campoleoni:2013iha} and matching the first quantum $1/k$ correction for $N=2$ \cite{Raeymaekers:2014kea}.
We note that in the current variables, the full effect of quantization was simply encoded in the fact that the quantum background charge
is shifted from its classical value according to (\ref{backgchargeshift}).\footnote{Note that in our setup, this should not be viewed as a quantum correction to the coupling constant $k$, since  $k$ plays a double role: it determines both the size of the winding mode lattice (\ref{a0quantcl}) and the value of the background charge, and the only the latter quantity receives quantum corrections according to (\ref{backgchargeshift}).}

\section{Discussion and outlook}\label{secdisc}
In this work we have formulated the $sl(N,\CC)$ higher-spin theories in the diagonal gauge,
and shown that this leads to a classical version of the Coulomb gas formalism. In particular, we saw that the screening charges emerge naturally as generators of residual gauge symmetries. As an example
of the power of this formulation, we completed the proof that the soliton-like winding solutions in the classical theory  represent  the $(0, \vec \L)$  degenerate primaries in the dual CFT. Before discussing some open questions, we now would like to comment further on the implications of the $\ZZ_2$ symmetry which we encountered in (\ref{Z2symm}).

\subsection*{Quantum duality and the matter spectrum}

In section~\ref{secquant} we showed that the observables in the quantum theory are invariant under the following symmetry:
\be
\a_+ \leftrightarrow \a_- \qquad {\rm or} \qquad \sqrt{k} \leftrightarrow - \frac{1}{\sqrt{k} }\, .\label{Z2}
\ee
This is a symmetry of the quantum $W_N$ algebra, since it leaves the Miura operator (\ref{qMiura}) invariant, which is not visible in the classical large $k$ limit \cite{Fateev:1987zh}.  It can be seen
as a part of the triality symmetry of the $\calw_\infty [N]$ algebra which survives the truncation to $W_N$
\cite{Gaberdiel:2012ku}.  

It is interesting to explore the consequences of  the following two natural assumptions:
\begin{itemize}
    \item (\ref{Z2}) is a symmetry of the full quantum theory, in particular it is also a symmetry of the spectrum. 
    \item the spectrum contains the soliton-like winding states (\ref{windings2}) with $\vec \a_0 =\a_- \vec \L$.
\end{itemize}
 The $\ZZ_2$ symmetry then predicts that the theory should also contain the states
with $\vec \a_0 =\a_+ \vec \L$, which are the primaries of the degenerate representation $(\vec \L,0)$.
These do not correspond to regular classical configurations of the higher-spin gauge field, since they do not satisfy the condition (\ref{a0quantcl})  and therefore the gauge field contains a singularity  corresponding to an external point particle  source  in the form of a Wilson line  \cite{Witten:1989sx,Ammon:2013hba}. The inclusion of these states therefore requires adding matter particles to the massless higher-spin theory. The energy of these matter states behaves at large $k$ as
\be
\D_2 (\a_+ \vec \L')\approx - \vec \L'\cdot \vec \r + \calo (k^{-1})\, .
\ee
The lightest of these excitations comes from taking
$\vec \L = \vec \e_1$ to be the highest weight of the ${\bf N}$-dimensional representation and
has $\D_2 = \bar \D_2= \frac{1-N}{2}$. This corresponds to a scalar particle with mass
\be
M^2 l^2 = 4 \D_2 (\D_2-1) = N^2-1\, .
\ee
This is precisely the mass of the scalar field coupled to $hs[\l ]$ massless higher-spin fields in the theory of Prokushkin and Vasiliev \cite{Prokushkin:1998bq}, in the
$\l \to N$ limit. One can also show that the full set of degenerate primaries $(\vec \L', \vec \L)$ arises from considering  scalar field excitations in soliton backgrounds \cite{Perlmutter:2012ds}.  Therefore the assumptions above essentially predict that the matter spectrum the theory agrees with that of the  Prokushkin-Vasiliev theory. 
 It would be interesting to generalize these results for the $SL(N,\CC)$ theory to the $hs[\l ]$ setting \cite{Campoleoni:2013lma}.

On a related note, one observes that the $\ZZ_2$ symmetry (\ref{Z2}) is somewhat reminiscent of T-duality, since it inverts $\sqrt{k}$ which sets the
scale for the winding lattice,  $\sqrt{k} W$, with $W$ the weight lattice. It is in general not quite T-duality however, since that would also replace the weight lattice by its dual, the root lattice $R$, divided by two.  Only for pure gravity, $N=2$, this is ordinary T-duality, since in that case we have that $R/2 = W$. The fact that the degenerate representations of the Virasoro algebra can be viewed as momentum and winding modes of a compact free boson was already pointed out in \cite{Felder:1988zp} (see also the early work \cite{Kadanoff:1978ve}). It is however amusing to note that, for general $N$, the  $\ZZ_2$ symmetry can be viewed as a T-duality in the theory of $N-1$ compact bosons in the presence of a B-field, as we show in Appendix \ref{appBfield}. However, the physical interpretation of such a B-field is not clear to us at present.

\subsection*{Outlook}
We conclude by pointing out some open questions and possible applications of our work.
\begin{itemize}
    \item We have dealt with the residual gauge symmetries of the diagonal gauge in  a somewhat ad hoc manner, which led us essentially to the original version of the Coulomb gas formalism in \cite{Fateev:1987zh}. It would be interesting to see if  a more rigorous treatment
    using BRST  methods would lead similarly to the BRST formulation of the Coulomb gas \cite{Felder:1988zp,Feigin:1990pn}.  It might be hoped that the connection to  bulk higher-spin symmetry would give a natural explanation for some  of the less obvious computational rules of the Coulomb gas formalism. 
\item We have mainly  focused here on classical solutions with simple poles or `centers' only in  the origin and at infinity, which
encode two-point functions of heavy operators in the dual  CFT. It would be interesting to explore if the diagonal gauge facilitates  the construction of multi-centered solutions which would encode  $n$-point correlators  of heavy operators.
\item It will also be interesting to include chemical potentials in the diagonal gauge in order to discuss higher-spin black holes \cite{GK_black-hole}. 
\item The free field parameterization discussed in this work may also prove useful in the context of the 
AGT correspondence \cite{Alday:2009aq,Wyllard:2009hg}. In that context,  as explained in \cite{Cordova:2016cmu},  a particular compactification of the $a_{N-1}$ (2,0) theory  gives rise to an $sl(N, \CC)$ Chern-Simons theory with Nahm pole boundary conditions which implement the Drinfeld-Sokolov constraint (\ref{asads}).
\end{itemize}

\acknowledgments

We thank Laszlo Feh\'er, Olaf Kr{\"u}ger and Tom\'a\v{s} Proch\'azka for useful discussions.
A.C.\ acknowledges the support of the Universit\'e libre de Bruxelles, where part of this work has been performed. His work has been partially supported by the ERC Advanced Grant ``€œHigh-Spin-Grav'', by FNRS-Belgium (convention FRFC PDR T.1025.14 and convention IISN 4.4503.15) and by the
NCCR SwissMAP, funded by the Swiss National Science Foundation.
The research of J.R.
was supported by the Grant Agency of the Czech Republic under the grant 17-22899S,
and by ESIF and MEYS (Project CoGraDS - CZ.02.1.01/0.0/0.0/15 003/0000437).
This collaboration was supported by the bilateral collaboration grant WBI 14-1 between the F\'ed\'eration Wallonie-Bruxelles and the Academy of Sciences of the Czech Republic. A.C. and J.R. wish to thank U.~Vienna for hospitality during the completion of this work.

\begin{appendix}

\section{Gravity subsector}\label{AppNewton}
Here we review how the theory (\ref{SE}) includes Euclidean AdS$_3$ gravity,
and compute the relation between $k$ and  Newton's  constant. For this we must choose a spin-2 subsector, which we take to correspond to the principal embedding of $sl(2,\CC)$ into $sl(N,\CC)$.
With this choice, it is convenient for the present purpose to take the  $sl(2,\CC)$ generators to be the $N$-dimensional  $su(2)$ representation
matrices $K_i$ $(i = 1,2,3)$ satisfying 
\bea
[K_i, K_j] &=& -\,\e_{ijk}\d^{kl} K_l\, ,\\
\tr_{\bf N} K_i K_j &=& - \frac{N(N^2-1)}{12}\, \d_{ij}\, ,\\
K_i^\dagger =&=& - K_i\, ,
\eea
with $\e_{123} =1$ (for example, for $N=2$ we can take $K_j = i \s_j/2$).
We restrict $A$ to lie in this $sl(2,\CC)$ subalgebra and decompose
\be
A = \left( \o^j + \frac{i}{l} e^j \right) K_j
\ee
with $\o^j$ and $e^j$ real one-forms. The $su(2)$-valued part $\o$ plays the role of the spin connection, while $e$ is the dreibein. Indeed, one shows that the Euclidean action (\ref{SE}) can, up to a boundary term, be written as
\bea
I_E &=& \frac{k}{l \p}  \int \tr_{\bf N} \left( e \wedge \cR - \frac{1}{3 l^2 } e\wedge e\wedge e\right)\\
&=& - \frac{N(N^2-1) k}{24 \p l }\int d^3 x \det e \left( R +\frac{2}{l^2}\right) ,\label{EH}
\eea
where $\cR= d\o + \o\wedge \o$ is the curvature  two-form, with  $\cR_i = \half \e_{ijk} \cR^{jk}$.
 Choosing the orientation such that $\det e >0$, we can identify $ \det e = \sqrt{g}$ and the action (\ref{EH}) takes the Einstein-Hilbert
 form. Therefore the theory (\ref{SE}) contains a spin-2 gravity sector with $k$ related to the AdS radius and Newton's constant as in (\ref{kitoGN}).

\section{Infinitesimal symmetries of Miura transform}\label{appMiura}
We will investigate here the infinitesimal symmetries of the classical Miura transformation~\eqref{classmiura}. Under an infinitesimal transformation 
\begin{equation}
    i\pa \vec\f \to i\pa \vec\f + \vec\eta \, ,
\end{equation}
the left hand side of the Miura transformation~\eqref{classmiura} changes as
\begin{align}
    &\big(-\sqrt{k}\big)\, \d M = \nonumber\\ & \sum_{a=1}^{N-1} \left( \prod_{j=1}^{a-1}\left( \pa +\frac{\vec\e_j \cdot (i\pa\vec\f)}{\sqrt{k}}\right) \right) \! \left( \pa\eta_a+\eta_a \frac{\vec e_a \cdot (i\pa\vec\f)}{\sqrt{k}} \right) \! \left( \prod_{j=a+2}^{N}\left( \pa +\frac{\vec\e_j \cdot (i\pa\vec\f)}{\sqrt{k}}\right) \right) ,
\end{align}
where we expanded $\vec\eta=\sum_{a=1}^{N-1}\eta_a \vec e_a$. The new combination that appears in the sum above will be denoted as
\begin{equation}\label{defeta1}
    \eta_{a,1}:= \pa\eta_a+\eta_a \frac{\vec e_a \cdot (i\pa\vec\f)}{\sqrt{k}} \, .
\end{equation}
In the expression for $\d M$ above we now move all factors containing a derivative to the right, and we obtain
\begin{align}
    \big(-\sqrt{k}\big)\,\d M & = \left(\eta_{1,1} + \dots + \eta_{N-1,1} \right)\prod_{j=3}^{N}\left( \pa +\frac{\vec\e_j \cdot (i\pa\vec\f)}{\sqrt{k}}\right)  \nonumber\\
     & \quad + \left( \eta_{2,2} + \dots + \eta_{N-1,2} \right) \prod_{j=4}^{N}\left( \pa +\frac{\vec\e_j \cdot (i\pa\vec\f)}{\sqrt{k}}\right)  \nonumber\\
     &\quad\ \,  \vdots \nonumber\\
     &\quad + \left( \eta_{N-2,N-2} + \eta_{N-1,N-2} \right)\left( \pa +\frac{\vec\e_N \cdot (i\pa\vec\f)}{\sqrt{k}}\right)\nonumber\\
     &\quad + \eta_{N-1,N-1} \, ,
 \end{align}
where we recursively introduced the combinations
\begin{equation}\label{defetaab}
   \eta_{a,b} := \left( \pa + \frac{i}{\sqrt{k}} (\vec\e_{a+1-b}-\vec\e_{a+1})\cdot (\pa\vec\f)\right) \sum_{l=a}^{N-1} \eta_{l,b-1} \, .
\end{equation}
Note that this is consistent with~\eqref{defeta1} when we identify $\eta_{a,0}:= \eta_a - \eta_{a+1}$ (setting $\eta_N =0$). 
Requiring that the left hand side of the Miura transformation does not change, $\d M=0$, then leads us to the condition
\begin{equation}\label{condeta}
    \sum_{a=b}^{N-1} \eta_{a,b} = 0 \,, \qquad b=1,\dots, N-1\, . 
\end{equation}
Any $\eta_a$ that satisfies~\eqref{condeta} with~\eqref{defetaab} defines an infinitesimal symmetry of the Miura transformation. We now show that such a transformation can always be obtained by a residual gauge transformation. For that purpose we define an upper triangular matrix $\tilde\l$ by
\begin{equation}\label{tildelambdafrometa}
    \tilde\l_{i,j} = e^{\frac{i}{\sqrt{k}}(\vec\e_i - \vec\e_j)\cdot \vec\f} \sum_{a=j-1}^{N-1} \eta_{a,j-i-1} \quad \text{for}\ j>i\, .
\end{equation}
We claim that this matrix satisfies equation~\eqref{eqlt}, and hence belongs to a transformation that preserves the diagonal gauge. To show this we note that the equations~\eqref{condeta} satisfied by the $\eta$'s allow us to formally extend~\eqref{tildelambdafrometa} to $i=0$ by setting $\tilde\l_{0,j}=0$. More trivially, the identification~\eqref{tildelambdafrometa} is also consistent with setting $\tilde\l_{i,N+1}=0$. Having said this, we can evaluate the right hand side of~\eqref{eqlt} for any $1\leq i <j\leq N$ using~\eqref{tildelambdafrometa} and we obtain:
\begin{align}
    e^{ - {\frac{i}{\sqrt{k}}\, \vec e_{i-1} \cdot \vec \f} } \tilde \l_{i-1,j} - e^{ - {\frac{i}{\sqrt{k}}\, \vec e_{j} \cdot \vec \f} } \tilde \l_{i,j+1} 
    &= e^{ - {\frac{i}{\sqrt{k}} (\vec\e_{i-1}-\vec\e_i  ) \cdot \vec \f} }
    e^{  {\frac{i}{\sqrt{k}} (\vec\e_{i-1}-\vec\e_j  ) \cdot \vec \f} } 
    \sum_{a=j-1}^{N-1} \eta_{a,j-i} \nonumber\\
    &\quad -  e^{ - {\frac{i}{\sqrt{k}} (\vec\e_{j}-\vec\e_{j+1}  ) \cdot \vec \f} }
    e^{  {\frac{i}{\sqrt{k}} (\vec\e_{i}-\vec\e_{j+1}  ) \cdot \vec \f} }
    \sum_{a=j}^{N-1} \eta_{a,j-i}\\
    &= e^{  {\frac{i}{\sqrt{k}} (\vec\e_{i}-\vec\e_{j}  ) \cdot \vec \f} } \,\eta_{j-1,j-i} \, .
\label{rhsofeqlt}
\end{align}
On the other hand, we plug~\eqref{tildelambdafrometa} into the left hand side of~\eqref{eqlt}, and we find
\begin{align}
    \partial \tilde\l_{i,j} &= e^{\frac{i}{\sqrt{k}}(\vec\e_i - \vec\e_j)\cdot \vec\f} \left( \pa + \frac{i}{\sqrt{k}}(\vec\e_{i}-\vec\e_{j})\cdot (\pa\vec\f)\right) \sum_{a=j-1}^{N-1} \eta_{a,j-i-1}\\
    &= e^{\frac{i}{\sqrt{k}}(\vec\e_i - \vec\e_j)\cdot \vec\f} \,\eta_{j-1,j-i}\, ,
\end{align}
which agrees with~\eqref{rhsofeqlt}. Hence, the upper triangular matrix $\tilde\l$ defines an allowed transformation. We can quickly check that its effect on $i\pa\vec\f$ is just a shift by $\vec\eta$ by observing that 
\begin{equation}
    \l_{i,i+1} = e^{  -{\frac{i}{\sqrt{k}} (\vec\e_{i}-\vec\e_{i+1}  ) \cdot \vec \f} } \,\tilde\l_{i,i+1} = 
    \eta_i\, .
\end{equation}
We conclude that the most general infinitesimal symmetry of the Miura transformation corresponds to a residual gauge symmetry in the diagonal gauge.

\section{Null vectors and their classical limits }\label{appsymm}

It is  instructive to see how our construction of classical null vectors in section \ref{secclassnull} gets corrected in the quantum theory \cite{Fateev:1987zh}. First we observe that the quantum-corrected version of the finite screening transformations (\ref{a0equiv}) reads
\be 
\vec \a_0 \sim w \cdot \vec \a_0 \equiv w( \vec \a_0 +\tilde\a_0 \vec \r) - \tilde \a_0 \vec \r 
\ee
for any Weyl reflection $w$. It can be shown \cite{Bouwknegt:1992wg} that these identifications leave the   $W_N$ charges invariant.  To construct the null descendants  of our winding states $|\a_- \vec \L, 0\rangle$ we apply the screening charges $\hat S_a$ to the $W_N$ primaries
\be 
|w_0 \cdot (\a_- \vec \L) + \a_+ \vec e_a, 0\rangle = : e^{ i (w_0 \cdot (\a_- \vec \L) + \a_+ \vec e_a )\cdot \hat{\vec \f} (0) }: |0,0\rangle
\ee 
with $w_0$ the Weyl reflection defined in (\ref{w0}). Note that the shift by  $ \a_+ \vec e_a$ vanishes in the classical large $k$ limit and was therefore not visible in the classical discussion in section \ref{secclassnull}.
Using the free field  OPE of $\hat{\vec \f} (z)$ we find
\begin{align} 
\hat S_a |w_0 \cdot (\a_- \vec \L) + \a_+ \vec e_a, 0\rangle = \hspace{-20pt}& \nonu
\frac{i k}{(\L^{N-1-a} +1)!} & \left( \pa_z^{\L^{N-1-a} +1} :e^{-i \a_+ \vec e_a \cdot \hat{\vec \f} (z)}   e^{ i (w_0 \cdot (\a_- \vec \L) + \a_+ \vec e_a)\cdot \hat{\vec \f} (0) }: \right)_{|z=0} |0,0\rangle\nonu 
\sim & \left(-  \frac{i \sqrt{k}}{\L^{N-1-a} +1}   \vec e_a \cdot \hat{\vec\a}_{-  (\L^{N-1-a} +1)} +\calo (1 )\right)  |w_0 \cdot (\a_- \vec \L), 0\rangle \, .
\end{align}
This state is a $W_N$ primary by construction and, from the  right-hand side, is also  a descendent  of $|w_0 \cdot (\a_- \vec \L), 0\rangle \sim |\a_- \vec \L, 0 \rangle $  at level $\L^{N-1-a} +1$. In the last line we displayed the leading large $k$ part, which agrees with the classical result (\ref{negnull}).

\section{All degenerate primaries as momentum-winding states}\label{appBfield}
We will show here that  $all$ $W_N$  degenerate primaries can be described as pure momentum-winding states in a version of the Coulomb gas formalism, where the free fields take values on a certain torus and  a constant $B$-field is turned on. The only effect of the latter is to modify the winding lattice, see e.g.\ \cite{Blumenhagen:2013fgp}.

Recall that, in the free field formalism with   $N-1$ free fields,  a general degenerate $W_N$-primary corresponds to a state $|\vec \a_0, 0\rangle$, where the zero-mode eigenvalue is
\be 
\vec \a_0 = \a_+ \vec \L'  + \a_- \vec \L \, . \label{a0quant}
\ee
Here, $\vec \L$ and $ \vec \L'$ are dominant weights of $sl(N)$, and $\a_\pm$ are related to the background charge as in (\ref{apm}).

We want to reproduce the zero mode quantization condition (\ref{a0quant}) as arising from momentum and winding number quantization 
in a theory where the free fields $\vec \f$ are  periodic, taking values in some torus, in the presence of  a constant   $B$-field. If the periods are 
\be
\vec \f \sim \vec \f + 2 \p\, n^a \vec l_a \qquad (\text{summation over}\ a = 1, \ldots, N-1)
\ee
where $n^a \in \ZZ$, $\{ l_a \}_a$ form a lattice basis, the left-moving zero mode is quantized as (see e.g.\  \cite{Blumenhagen:2013fgp}, eq. (10.49), with $\a'=2$)):
\be
\vec \a_0 = m_a \vec l^{*a} - \half w^a (g_{ab} + b_{ab} )\vec l^{*b}  \label{pL}
\ee
where $m_a$ and  $w^a$ are  integer momentum and winding numbers,  $\{ l^{*a} \}_a$ are dual basis vectors satisfying $ l_a \cdot   l^{*b} = \d_a^b$, $g_{ab} =l_a \cdot l_b$, and $b_{ab}$ is the constant $B$-field.

We now take the period lattice to be the root lattice divided by $\a_+$, with  basis vectors  given by
\be 
\vec l_a = \frac{1}{\a_+ }\, \vec e_a\, ,
\ee 
so that the dual lattice is the weight lattice multiplied by $\a_+$, with dual basis vectors
\be 
\vec l^{*a} = {\a_+ }\vec \o_a\, .
\ee
The metric $g_{ij}$ is proportional to the $sl(N)$ Cartan matrix,
\be 
g_{ij}= \frac{1}{\a_+^2 }\, C_{ij}=  \frac{1}{\a_+^2 }  \left( \begin{array}{cccc} 2 & -1 & 0 & \ldots \\-1& 2& -1& \ldots\\ 0&-1& 2 &-1\\ &&& \ldots \end{array}\right)\, .
\ee
Finally, for the $B$-field we take
\be 
b_{ij} = \frac{1}{\a_+^2 }  \left( \begin{array}{cccc} 0 & 1 & 0 & \ldots \\-1& 0& 1& \ldots\\ 0&-1&  &0\\ &&& \ldots \end{array}\right)\, .
\ee
Plugging this all into (\ref{pL}), we obtain indeed (\ref{a0quant}) with
\bea
\vec \L &=& m_a \vec \o_a \\
\vec \L' &=& w^a \vec \e_a
\eea
where the $\vec \e_a$ are, as before, the first $N-1$ weights of the fundamental representation,
\be \vec \e_a = \vec \o_a - \vec \o_{a-1} \, . \label{transfapp} \ee
The $\vec \e_a$ also form a basis of the weight lattice; this follows from the fact that the transformation (\ref{transfapp})  involves integer coefficients and so does its inverse
\be
\vec \o_a = \sum_{b=1}^a \vec \e_b\, .
\ee
Hence, by choosing the momentum and winding numbers  such that  $\vec \L$ and $ \vec \L'$ are dominant weights, all the degenerate representations (\ref{a0quant}) are reproduced as momentum-winding
states in this theory. We note that for $N=2$, where the $B$-field is absent, this reduces to the observation
 made in \cite{Felder:1988zp}.

\end{appendix}

\end{document}